\documentclass[preprint, aps, pra, footinbib, letterpaper, superscriptaddress]{revtex4-1}

\usepackage[T1]{fontenc} 
\usepackage{amsmath,color}
\usepackage{amssymb}
\usepackage{amsfonts}
\usepackage{amsthm}
\usepackage{bm}
\usepackage{mathtools}
\usepackage[title]{appendix}
\usepackage{multirow}

\newcommand*{\citen}[1]{%
  \begingroup
    \romannumeral-`\x 
    \setcitestyle{numbers}%
    \cite{#1}%
  \endgroup
}

\usepackage{algorithm}
\usepackage{algpseudocode}
\usepackage{dsfont}

\usepackage{bbold}

\DeclarePairedDelimiterX\braket[2]{\langle}{\rangle}{#1 \delimsize\vert #2}

\begin{document}

\title{Heat Conduction in Polymer Chains: Effect of Substrate on the Thermal Conductance}

\author{Mohammadhasan Dinpajooh$^{1}$\footnote{Corresponding author. Electronic mail: mdinpajo@sas.upenn.edu} and Abraham Nitzan}
\affiliation{Department of Chemistry, University of Pennsylvania, Philadelphia, Pennsylvania 19104, USA }
\affiliation{School of Chemistry, Tel Aviv University, Tel Aviv 69978, Israel}

\date{\today}

\begin{abstract}
\vspace{1 cm}
In standard molecular junctions, a molecular structure is placed between and connected to metal leads. Understanding how mechanical tuning in such molecular junctions can change heat conductance has interesting applications in nanoscale energy transport. In this work, we use nonequilibrium molecular dynamics simulations to address the effect of stretching on the phononic contribution to the heat conduction of molecular junctions consisting of single long-chain alkanes and various metal leads such as Ag, Au, Cu, Ni, and Pt. 
The thermal conductance of such junctions is found to be much smaller than the intrinsic thermal conductance of the polymer and significantly depends on the nature of metal leads as expressed by the metal-molecule coupling and metal vibrational density of states. This behavior is expected and reflects the mismatch of phonon spectra at the metal molecule interfaces. As a function of stretching, we find a behavior similar to what was observed earlier [J. Chem. Phys. {\bf 153}, 164903 (2020)] for pure polymeric structures.
At relatively short electrode distances, where the polyethylene chains are compressed, it is found that the thermal conductances of the molecular junctions remain almost constant as one stretches the polymer chains. At critical electrode distances, the thermal conductances start to increase, reaching the values of the fully-extended molecular junctions. 
Similar behaviors are observed for junctions in which several long-chain alkanes are sandwiched between various metal leads. 
 These findings indicate that this behavior under stretching is an intrinsic property of the polymer chain and not significantly associated with the interfacial structures. 
\end{abstract}
 \maketitle
		

\section{Introduction}
\label{Intro}

Controlling the thermal properties of molecular devices such as molecular junctions can support technologies based on heat management at the nanoscale.\cite{Cui2017,Wang2020}
The thermal conductance measurements in molecular junctions involve molecular structures sandwiched between conducting substrates. Such substrates are usually metals in which the heat is carried primarily by electrons while the heat transport in the molecular structures is usually dominated by nuclear motions.
The role of electron transport in molecular heat conduction should not, however, be disregarded.\cite{Huberman1994,Majumdar2004,Mahan2009,Lombard2015,Sadasivam2015,Giri2020}
The electron-phonon coupling has been discussed to be considerable for the metal/semiconductor interfaces such as titanium silicide/silicon interface\cite{Sadasivam2015} and metal/ionic interfaces, where the surface charges may provide a matrix element between the metal electrons and the phonons in the insulator.\cite{Mahan2009}
Electron transfer induced heat transfer has been suggested as a heat transport mechanism in molecular junctions and Pauly {\it et al.} suggested that the electronic contribution to the thermal conductance can be significant for molecular junctions consisting of two CH$_2$ units.\cite{Klockner2016} 
However, here we focus on the phononic heat transport mechanism that is expected to dominate heat conduction in junctions comprising long alkane chains. We follow previous studies of the possibilities to exert external control through mechanical forces,\cite{Li2015,Vacek2015,Dulic2003,Cui2019} and focus on a similar model in the present study.

Recent studies suggested very high thermal conductances for stretched polymer chains\cite{Liu2010,Liu2012,Chen2020a} suggesting that they can be used as heat exchangers. In a previous work, we investigated the thermal conductances of various polymer chains upon stretching in the absence of metal leads, i.e. intrinsic thermal conductances of polymer chains.\cite{Dinpajooh2020} We showed that the nature of heat transport along such chains was a threshold phenomenon: at relatively small end-to-end distances the thermal conductances remain almost constant as one stretches the polymer chain, while at critical end-to-end distances thermal conductances start to increase, reaching the fully-extended chain values. In addition, consistent with previous studies,\cite{Zhang2012} we also found a similar threshold behavior for aligned crystalline fibers.\cite{Dinpajooh2020}
Here, we extend our studies to polymer chains connecting metal substrates. 

A significant amount of work has already been performed to understand heat conduction in molecular junctions consisting of relatively small molecules. Experimentally, complications can arise due to the uncertainty in the number of molecules in contact with the electrodes because of roughness and not-chemically-bound molecules to both electrode surfaces.\cite{Mosso2019} Additionally, residues and spurious molecules can be present on the surface and the target molecules might not directly bridge the electrodes. Therefore, interpreting observations of mechanical control of heat conduction can be assisted by detailed numerical simulations. 

Here, we use molecular dynamics simulations to address the magnitudes of geometry-induced variations of the thermal conductances for simple molecular junctions consisting of reasonably large molecules and ask how the molecular junction thermal conductances change as the electrode distances change. 
In particular, we study the single dithiolated large alkane molecules (polyethylene) and investigate the heat conduction in various metal-polyethylene-metal junctions, where the polyethylene polymer chains are attached to metal substrates by sulfur atoms. The metal substrates consist of Ag, Au, Cu, Ni, and Pt. We change the electrode distances and report the corresponding thermal conductances for these junctions. In particular, we (i) examine the effect of various metal leads on the thermal conductance and the correlation between the substrate spectrum and molecular vibrational density of states, (ii) address the influence of the metal-molecule coupling and possible anharmonic effects, (iii) study the resulting effect of the structural parameters such as torsional gauche defects upon stretching, and (iv) compare the thermal conductances of single molecular junctions with the nanowire junctions consisting of several polymer chains. We analyze these behaviors with classical nonequilibrium molecular dynamics simulations at room temperature, disregarding possible electronic contributions to the heat conductance assuming such contributions are not very significant. One of the advantages of nonequilibrium molecular dynamics simulations is that no assumptions are made about the nature of scattering events occurring at the interface, but one should note the limits of classical molecular dynamics simulations when performed close to the Debye temperatures of the materials of interest.\cite{Landry2009,MoghaddasiFereidani2019}

The remainder of the paper is organized as follows: in the
next section, the numerical/simulation details are summarized. In
Sec. \ref{results}, the results of simulations are discussed, and 
Sec. \ref{conclusion} summarizes the main conclusions.

\section{Numerical Details}
\label{details}

The molecular dynamics (MD) software program LAMMPS\cite{Plimpton1995} (a corrected version by Boone {\it et al.}\cite{Boone2019}) was used for classical nonequilibrium molecular dynamics (NEMD) simulations to obtain the thermal conductance in this study. The molecular junction was created by sandwiching the polyethylene (PE) chain molecules between two metal leads. The outermost metal atoms were fixed such that one can set the electrode distance at a given distance.
The temperature of all other metal atoms except those in the four innermost layers next to the polymer were set at the imposed left and right temperatures using white (Markovian) Langevin thermostats. The target temperatures for these hot and cold regions were $320$ and $280$ K, respectively.
Following relaxation to steady state (about 4 ns), the NEMD simulations were continued for $40$ ns in order to collect data for statistical averaging. The system was propagated in time with a timestep of $0.25$ fs. At the steady state, the energy flux can be computed as the energy per unit time taken out from the cold end or injected into the hot end. Specifically, we use 

\begin{equation}
I_z  = \frac{1}{2}\left(  \left|\frac{\Delta E_{\rm{hot}} }{\Delta t}\right|  + \left|\frac{\Delta E_{\rm{cold}} }{\Delta t}\right|     \right), 
\label{IzNEMD}
\end{equation}
where at the steady state $\Delta E_{\rm{hot}}$ and $\Delta E_{\rm{cold}}$ are the amounts of energy added to the hot region and subtracted from the cold region during a time interval $\Delta t$ to create the heat flux.
A thorough explanation of these calculations can be found in Ref \citen{Boone2019}.

\begin{figure*}
\includegraphics[width=\columnwidth]{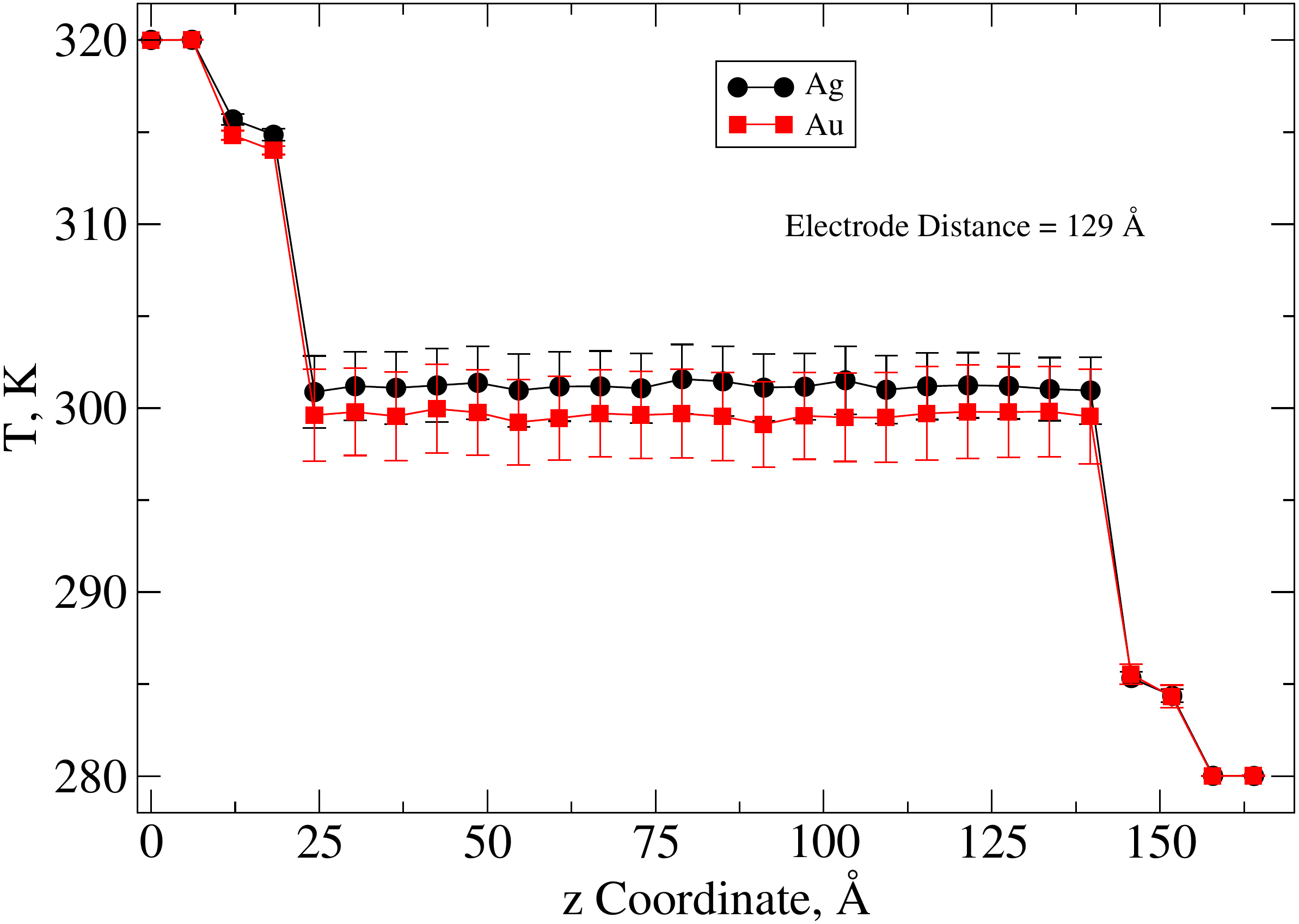}
\caption{
The steady state temperature profiles for gold-polyethylene-gold junction and silver-polyethylene-silver junction when the left and right temperatures are set to $T_L=320$ K and $T_R=280$ K. Note that the Langevin thermostat is applied only beyond the innermost four metal layers next to the molecule. The electrode distance for these molecular junctions is about $129$ \AA\ and the average temperature profiles and the error bars were obtained from ten independent nonequilibrium molecular dynamics simulations.
}
\label{figTprof}
\end{figure*}

Setting the direction perpendicular to the metal leads as $z$, the temperature profile is obtained by monitoring the kinetic energy of chain atoms within (typically $30$) slabs of equal size, $\Delta z$ along the $z$ axis. An example of the computed temperature profile for a gold-polyethylene-gold junction is seen in Fig. \ref{figTprof}.

In this report, the overall thermal conductance, $K$, was obtained as $K = I_z/\Delta T$, where $I_z$  is the heat current and $\Delta T$ is the average temperature difference between the hot and cold Langevin metal layers. Note that there is a temperature drop/difference at each metal/polymer interface. Standard velocity-Verlet time integrator was used and no periodic boundary conditions were used in the $z$ direction for these NEMD simulations. 

As briefly described above, the system studied in this work comprises the polymer chains placed between and attached to metal substrates. 
The initial structure of each metal substrate (lead) was arranged in an fcc crystal structure and consisted of $500$ or $864$ metal atoms (unless mentioned otherwise). Consistent with the experimental values, the lattice parameters to create the initial fcc crystal structures of Ag, Au, Cu, Ni, and Pt metals were $4.09$, $4.08$, $3.62$, $3.52$, and $3.92$ \AA, respectively. The embedded atom method, which uses semi-empirical, many-atom potentials for computing the total energy of metallic systems, was used to treat metal-metal interactions in the metal leads.\cite{Daw1984,Foiles1986} Note that this method is related to the second moment approximation to tight binding theory.\cite{Ackland1988}

The initial structure for a PE polymer chain was obtained from the PE crystal structure. The TraPPE United Atom (UA) models are used to treat the intramolecular/intermolecular polymer interactions, which coarse-grain each $\mathrm{CH}_x$ unit to one interaction site\cite{Martin1998,Wick2000} to represent the polymeric units.
In this model, we assume that the stiff,
high frequency C$-$H bonds do not play a significant role in the thermal transport and
a harmonic potential is used to model two-body CH$_2$$-$CH$_2$ bond potentials. The TraPPE force field (FF) also consists of angle (three-body) potentials and dihedral (four-body) potentials, where the details are presented in Table \ref{param}. Therefore, the Hamiltonian of the polymer consisting of $N$ beads (for the TraPPE FF) is given by:

\begin{equation}
\begin{split}
H_{\rm{molecule}} & = \sum_i \frac{p_i^2}{2m_i} +  \sum_i k_{bi} (l_i-l_{0i})^2 + \sum_i k_{\theta i} (\theta_i-\theta_{0i})^2 + \\
  & \sum_i \sum_{n_i}^4 \frac{C_{n i}}{2} \left[ 1+ (-1)^{n{_i}-1} {\rm{cos}} ( n_i \phi_i) \right] + \sum_i \sum_j 4 \epsilon_{ij} \left[  (\frac{\sigma_{ij}}{r_{ij}})^{12} - (\frac{\sigma_{ij}}{r_{ij}})^{6}  \right],
\end{split}
\label{Hmol}
\end{equation}
where $p_i$ is the momentum of a given bead with mass of $m_i$, $l_i$ is the bond length between two given beads, $l_{0i}$ is the equilibrium bond length between two given beads, and $k_{bi}$ is the force constant for the harmonic potential of the bonds.
Similarly,  $\theta_i$ is the bending angle between three given beads, $\theta_{0i}$ is the equilibrium bending angle between three given beads, and $k_{\theta i}$ is the related force constant to control bond-angle bending.
The torsional potentials are used to restrict the dihedral rotations around bonds connecting two beads and consist of the dihedral coefficients $C_{ni}$ and the related angles $\phi_i$.  
In addition, the bead pairs separated by four bonds interact with each other by the Lennard-Jones interactions, which involve CH$_2$/CH$_2$ and CH$_2$/S interactions in a given polymer. A cutoff distance of $14$ \AA\ is used to truncate the Lennard-Jones interactions.

\begin{table}[tbh]
\centering
\caption{Interaction parameters for the thiolated polyethylene polymer chains using a United Atom force field\cite{Martin1998,Wick2000}. Note that the Lorentz-Berthelot rule is used for non-bonded LJ potentials.}
\begin{tabular}{ccc}
     \hline
     \multicolumn{2}{c}{Bond potential: $U_{\rm{bond}} =  k_b (l-l_0)^2$ }  \\
       & $k_b$, kcal mol$^{-1}$ \AA$^{-2}$ & $l_0$, \AA \\
      CH$_2$--CH$_2$  & $450$ &  $1.54$ \\
      CH$_2$--S           & $518$ &  $1.82$ \\
     \hline 
     \multicolumn{2}{c}{Angle potential: $U_{\rm{angle}} =  k_\theta (\theta-\theta_0)^2$ }  \\
       & $k_\theta$, kcal mol$^{-1}$ rad$^{-2}$ & $\theta_0$, deg \\
     CH$_2$--CH$_2$--CH$_2$  & $62.1$ & $114.0$ \\
     CH$_2$--CH$_2$--S          & $62.1$ & $114.4$ \\
     \hline 
     \multicolumn{2}{c}{Dihedral potential: $U_{\rm{dih}} =  \Sigma_n^4 \frac{C_n}{2} \left[ 1+ (-1)^{n-1} {\rm{cos}} ( n \phi) \right]$ }  \\
       & $C_i$, kcal mol$^{-1}$ & $C_i$ \\
      & $1.4110$  & $C_1$ \\
     CH$_2$--CH$_2$--CH$_2$--CH$_2$/S  & $-0.2708$ & $C_2$ \\
      & $3.1430$   & $C_3$ \\
       & $0$   & $C_4$ \\ 
     \hline
     \multicolumn{2}{c}{Non-bonded potential: $U_{\rm{LJ}} =  4\epsilon \left[ (\sigma/r)^{12} - (\sigma/r)^6 \right] $ }  \\
       & $\epsilon$, kcal mol$^{-1}$ & $\sigma$, \AA \\
      CH$_2$(sp$^3$) & $0.0912$ & $3.95$ \\
      S                      & $0.274$ & $3.59$ \\
     \hline
\end{tabular}
\label{param}
\end{table}

The PE polymer chains were attached to the metal leads by sulfur atoms and the bonding character of molecular metal-thiolate was described by the Morse potential, $D_e[\exp(-2\alpha(r-r_e))-2\exp(-\alpha(r-r_e))]$, between sulfur atoms and metal atoms.\cite{Jiang2002,Liu1999} The Morse metal-sulfur parameters are reported in Table \ref{paramMorse}. These Morse metal-sulfur parameters were obtained by making use of the appropriate Morse combination rules\cite{Yang2019} considering the Morse parameters for metals\cite{Pamuk1976} and sulfur bonds.
Finally, the metal-CH$_2$ interactions were treated by the Lennard-Jones potential with the $\epsilon_{\rm{M-CH_2}}$ and $\sigma_{\rm{M-CH_2}}$ obtained from the Universal Force Field\cite{Rappe1992} and are reported in Table \ref{paramMorse}.

\begin{table}[tbh]
\centering
\caption{Metal-polymer interactions: the Morse potential is used to model metal-sulfur (M-S) interactions while metal-CH$_2$ (M-CH$_2$) interactions are modeled using the LJ potential.}
\begin{tabular}{c|ccc|cc}
     \hline
      Metal (M) & \multicolumn{3}{c|}{Morse Potential: M-S} &  \multicolumn{2}{c}{LJ Potential: M-CH$_2$} \\
      \hline
      & $D_e$/eV$\;\;$ & $r_e$/\AA$\;\;$ & $\alpha$/\AA$^{-1}$$\;\;$ & $\epsilon$/eV & $\sigma$/\AA  \\
      \hline
     Ag      & $0.619$   &  $2.73$      & $1.38$      &  $0.0025$ & $3.38$   \\
     Au      & $0.879$   &  $2.70$      & $1.47$      &  $0.0026$ & $3.44$   \\
     Cu      & $0.652$   &  $2.64$      & $1.41$      &  $0.0009$ & $3.53$   \\
     Ni       & $0.812$   &  $2.61$      & $1.44$      &  $0.0016$ & $3.24$   \\
     Pt       & $1.239$   &  $2.67$      & $1.47$      &  $0.0037$ & $3.20$     \\
     \hline
\end{tabular}
\label{paramMorse}
\end{table}

The vibrational density of states for the PE polymer chains and metals were calculated in order to address the significance of vibrational overlap in controlling thermal conductance.
The vibrational density of states for the PE polymer chains connected to the electrodes were obtained from $10000$ configurations after minimizing each configuration, where the normal modes of these minimized structures were obtained by the diagonalization of the Hessian matrix.
The energies of polymer structures were minimized using a simple “steepest descent minimizer” such that the minimization was converged when the maximum force was smaller than $25$  kJ mol$^{-1}$  nm$^{-1}$. 
The vibrational densities of states for metal layers were obtained by Fourier transforming the velocity autocorrelations of an ensemble of
atoms in the four non-Langevin metal layers positioned next to the PE polymer chains (both left and right four metal layers that do not obey the Langevin dynamics). The vibrational density of states are obtained from

\begin{equation}
g(\omega) = \int C_{\mathbf{v}}(t) e^{-i \omega t} dt,
\label{dos_eq}
\end{equation}
where $C_{\mathbf{v}}(t)$ is the velocity correlation function calculated from MD trajectories according to 

\begin{equation}
\label{vat}
C_{\mathbf{v}}(t) = \frac{1}{n_{\rm{M}}} \sum_{i=1}^{n_{\rm{M}}} \frac {\langle \mathbf{v}_i(t+t_0) \cdot  \mathbf{v}_i(t_0) \rangle}{\langle \mathbf{v}_i(t_0) \cdot  \mathbf{v}_i(t_0) \rangle} ,
\end{equation}

where $n_{\rm{M}}$, $\mathbf{v_i}$, $t_0$, and $\tau$ are the number of atoms in four metal layers next to the polymer, vector velocity of atom $i$ in the aforementioned metal layers, starting time, and the autocorrelation time, respectively. We note that the vibrational density of states obtained for four non-Langevin metal layers at either the left or right side were not significantly different (results not shown).

\section{Results}
\label{results}

{\it Thermal Conduction of Different Junctions.}

\begin{figure*}[tbh]
\includegraphics[width=0.8\linewidth]{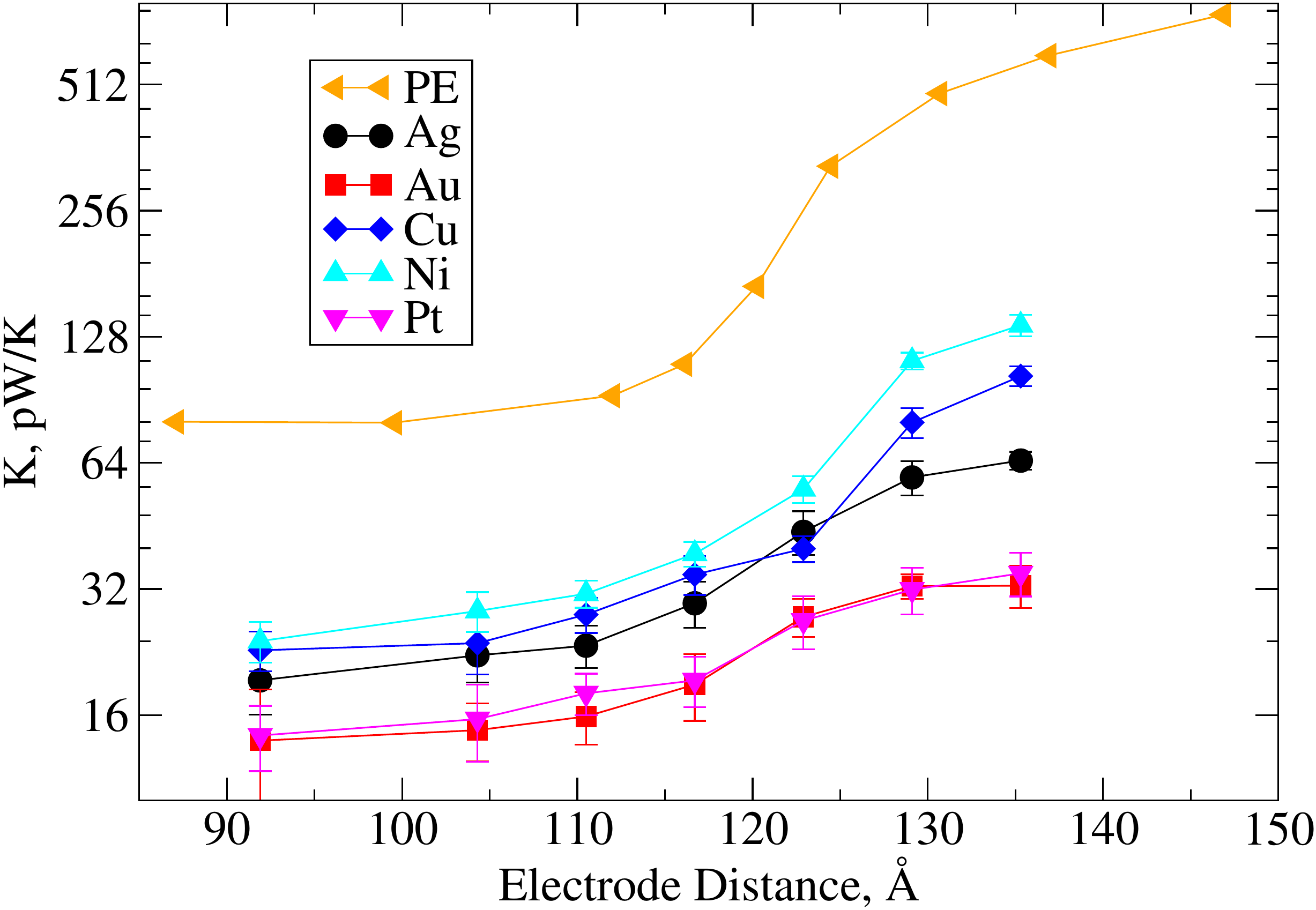}
\caption{
The effect of metal substrate on the thermal conductance of junctions at various electrode distances. The polyethylene (PE) chain, consisting of $96$ carbon atoms, is attached to metal substrates by the sulfur atoms. Note that the junction breaks at large stretching values of interelectrode distances, hence the absence of data beyond $\sim$ 135 \AA.
The intrinsic thermal conductance of the PE polymer chain is also shown in orange for comparison. 
}
\label{KM}
\end{figure*}

The thermal resistance of interfaces can be affected by the properties of the materials on each side of the interface, the atomic-level details of the interfacial structure, and the stiffness of interfacial bonds. Fig. \ref{KM} shows the thermal conductances at various electrode distances for the PE polymer chain consisting of $96$ carbons attached to metal substrates by sulfur atoms. 
Here, ten independent NEMD simulations have been performed to report statistical uncertainties.
As can be seen, heat transport in such realistic heterogeneous junctions is sensitive to the metal leads, and the metal interfaces limit the thermal conductance of molecule junctions. Comparing the orange line with other lines in Fig. \ref{KM}, it is clear that the thermal conductance increases dramatically when the metal substrate is replaced by a covalently connected PE chain. For relatively large values of the interelectrode distance, the molecular junctions break and the thermal conductances are not reported. The threshold phenomenon reported earlier\cite{Dinpajooh2020} for homogeneous systems in the behavior of heat conduction upon molecular stretching appears to persist also in the heterogeneous junctions. We may conclude that this behavior, transition from the compressed to stretched value of the heat conduction at a relative narrow range of the end-to-end distance, is an intrinsic property of the polymer chain while the absolute conductance strongly depends on the properties of molecule-metal interfaces. 

\clearpage

{\it Correlation with Structural Parameters.}

\begin{figure*}[tbh]
\includegraphics[width=0.8\linewidth]{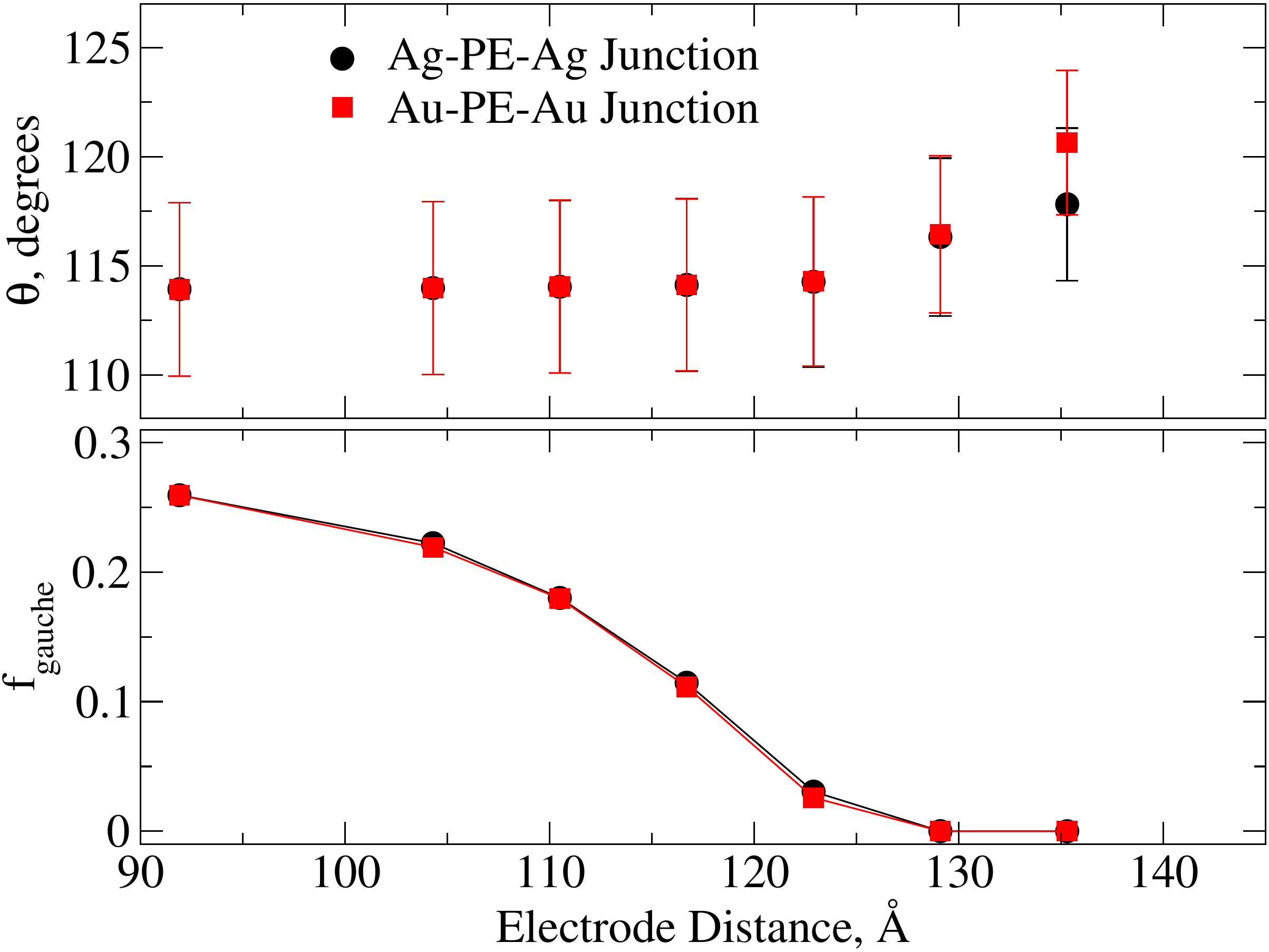}
\caption{
Top panel: The average and variance (displayed via the “error” bars) of the CCC
angular distribution as a function of electrode distance for a single polyethylene (PE) polymer chain sandwiched between gold and silver metals. 
Bottom panel: Fraction of gauche states as a function of electrode distance for the aforementioned PE polymer chain.
}
\label{CCC}
\end{figure*}

The average CCC bending angle as well as the fraction of gauche states (as previously defined in the supplementary information of Ref. \cite{Dinpajooh2020}) at various electrode distances are shown in Fig. \ref{CCC} for the PE polymer chains sandwiched between the aforementioned metal leads. As can be seen, both structural parameters depend strongly on stretching; however, it is the CCC bending angle that directly correlates with the heat conduction behavior. The average CCC bending angle is around $114$ degrees for the compressed polymer chains, and remains nearly constant until the electrode distances significantly increase. Beyond a threshold length, the CCC bending angle increases and reaches about $120$ degrees for stretched polymer chains. This threshold behavior of the CCC angle under stretching is similar to the heat conduction behavior. On the other hand, the fraction of gauche states decreases gradually as the electrode distance increases until it reaches zero at relatively large electrode distances and shows less correlations with the CCC angle threshold behavior.

\clearpage

{\it Density of States.}

\begin{figure*}[h]
\includegraphics[width=0.8\linewidth]{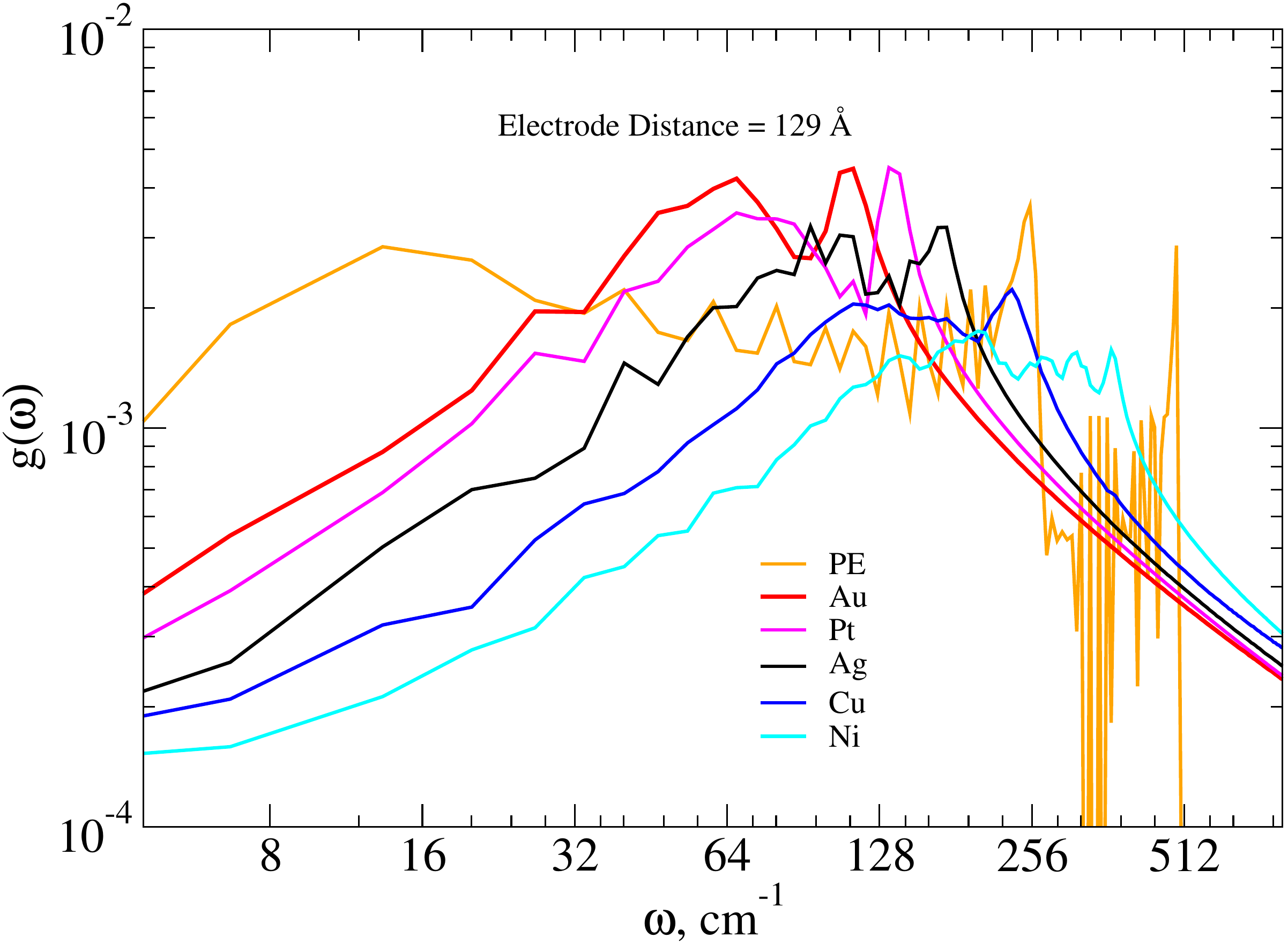}
\caption{
Comparison of the normalized vibrational density of states, $g(\omega)$ (Eq. \ref{dos_eq}), for various metal leads and the polyethylene polymer chain with 96 carbon atoms sandwiched between metal leads through sulfur atoms.
}
\label{DOSM}
\end{figure*}

Figure \ref{DOSM} shows the vibrational density of states for the metals investigated in this work along with the vibrational density of states of the PE polymer chain at a given interelectrode distance (namely stretching state). Since the low frequency vibrational modes dominate the room temperature heat transfer, Fig. \ref{DOSM} focuses on the low frequency domain. As can be seen, the density of states for the PE polymer chain increases initially up to about $16$ cm$^{-1}$, which goes to zero around $512$ cm$^{-1}$ showing a maximum around $260$ cm$^{-1}$. A mode localization analysis using harmonic force fields shows that the modes around $260$ cm$^{-1}$ are more localized than lower/higher frequencies (results not shown). On the other hand, the densities of states of the metals show maxima at different frequencies ranging from about $64$ cm$^{-1}$ to $200$ cm$^{-1}$. The density of states for the metals are obtained from the velocity autocorrelation functions, Eq. \ref{vat}, calculated for the metal atoms in the four metal layers next to the polymer as described in Section II. These atoms interact with their neighbor and the PE chain, but are not subjected to the Langevin forces and associated damping used to thermalize the system at the edge of its metal components. 

The different values of the thermal conductances of junctions comprising different metals can be rationalized by considering in the interfacial resistance. 
In the harmonic limit, the heat transfer occurs mostly elastically\cite{Hopkins2011,Saaskilahti2014} and the frequencies of the reflected and/or
transmitted phonons are the same as that of the incident phonon. In this limit, the interfacial conductance depends on the overlap of the density of states from each side of the interface given by\cite{Shen2011} 

\begin{equation}
\label{ovrlap}
O_{\rm{\alpha/\beta}} = \frac{|\int g_{\alpha}(\omega) g_{\beta}(\omega) d\omega|^2}{
\int g^2_{\alpha}(\omega)  d\omega
\int g^2_{\beta}(\omega) d\omega},
\end{equation}
where $g_{\alpha}(\omega)$ and $g_{\beta}(\omega)$ are the vibrational density of states for $\alpha$ and $\beta$ materials, respectively.\cite{Lu2020}
Note that there are alternative ways to quantify the vibrational overlaps but the definition in Eq. \ref{ovrlap} allows one to compare them such that $O_{\rm{\alpha/\beta}}$ goes to $1$ when the molecular chain and the electrode are made of the same material and vanishes when $g_{\alpha}(\omega) g_{\beta}(\omega) =0$. Table \ref{oval} shows the magnitudes of the vibrational overlaps between various metal leads and the PE polymer chain or the sulfur atom of the polymer chain.

\begin{table}[tbh]
\centering
\caption{The magnitudes of vibrational overlaps between various metal substrates (M) and the polyethylene polymer chain (PE) or the sulfur atoms of the polymer chain (S) at an electrode distance of $129$ \AA. Note that the density of states for metal substrates are obtained by considering the four non-Langevin metal layers next to the PE polymer chain. The last column in the Table lists the values of the parameter $\xi$ = $(K_{\mathrm{H}}-K_{\mathrm{L}})/(K_{\mathrm{H}}+K_{\mathrm{L}})$, where $K_{\mathrm{H}}$ and $K_{\mathrm{L}}$ are the thermal conductances of the molecular junctions at the largest and smallest electrode distances investigated, respectively.}
\begin{tabular}{ccccc}
     \hline
     Metal & $O_{\rm{M/PE}}$ & $O_{\rm{M/S}}$ & $\xi$ \\
     \hline
      Au     & $0.56$ &  $0.17$   &  $0.40$ \\
      Pt      & $0.59$ &  $0.17$   &  $0.42$ \\
      Ag     & $0.63$ &  $0.35$   &  $0.54$ \\
      Cu     & $0.72$ &  $0.43$   &  $0.64$ \\
      Ni      & $0.83$ &  $0.49$   &  $0.70$ \\
     \hline
\end{tabular}
\label{oval}
\end{table}

Table \ref{oval} together with Fig. \ref{KM} show that there is a good correlation between the thermal conductances of molecular junctions and the vibrational overlap between the PE polymer chain and metal substrates.
In addition, focusing on the vibrational density of states computed from the motion of the sulfur atoms next to the metal leads, the third column in Table \ref{oval} shows that a good correlation is obtained between the thermal conductance and vibrational overlap of metal leads and the sulfur atoms. The last column in Table \ref{oval} quantifies an appropriate way to address the effect of substrate on the thermal conductance upon stretching of the molecular junctions by defining $\xi = (K_{\mathrm{H}}-K_{\mathrm{L}})/(K_{\mathrm{H}}+K_{\mathrm{L}})$, where $K_{\mathrm{H}}$ and $K_{\mathrm{L}}$ are the thermal conductances corresponding to the largest and smallest electrode distances investigated. It is interesting to note that the stretching effect on the thermal conductance is larger for junctions with stronger interfacial overlap. Still, the threshold behavior in the conduction/stretching dependence appears in the studied molecular junctions, consistent with our assertion that this is an intrinsic molecular phenomenon.  

Other factors such as bond strength, stiffness, phonon coupling, and inelastic scattering can also influence the thermal conductance values.\cite{OBrien2013,Hopkins2011,Park2020}  
Inelastic scattering is included in our MD simulations because the atomic interactions are anharmonic. At the metal/molecule interface, anharmonic channels may open pathways for thermal transport across the junction that are not present in the harmonic channels.\cite{Luo2010,Luo2011a,Saaskilahti2014,Reid2019}
The correlations between the vibrational overlap and the thermal conductance observed in our calculation indicate that the heat transfer mechanism at the interface is mostly affected by the vibrational mismatches of the materials/molecules\cite{MoghaddasiFereidani2019,Majumdar2015,Luo2011a} and anharmonic channels do not dominate the thermal conductances in the molecular junctions studied in this work.

\vspace{0.25in}
{\it Metal-molecule Coupling.}

\begin{figure*}[tbh]
\includegraphics[width=0.8\linewidth]{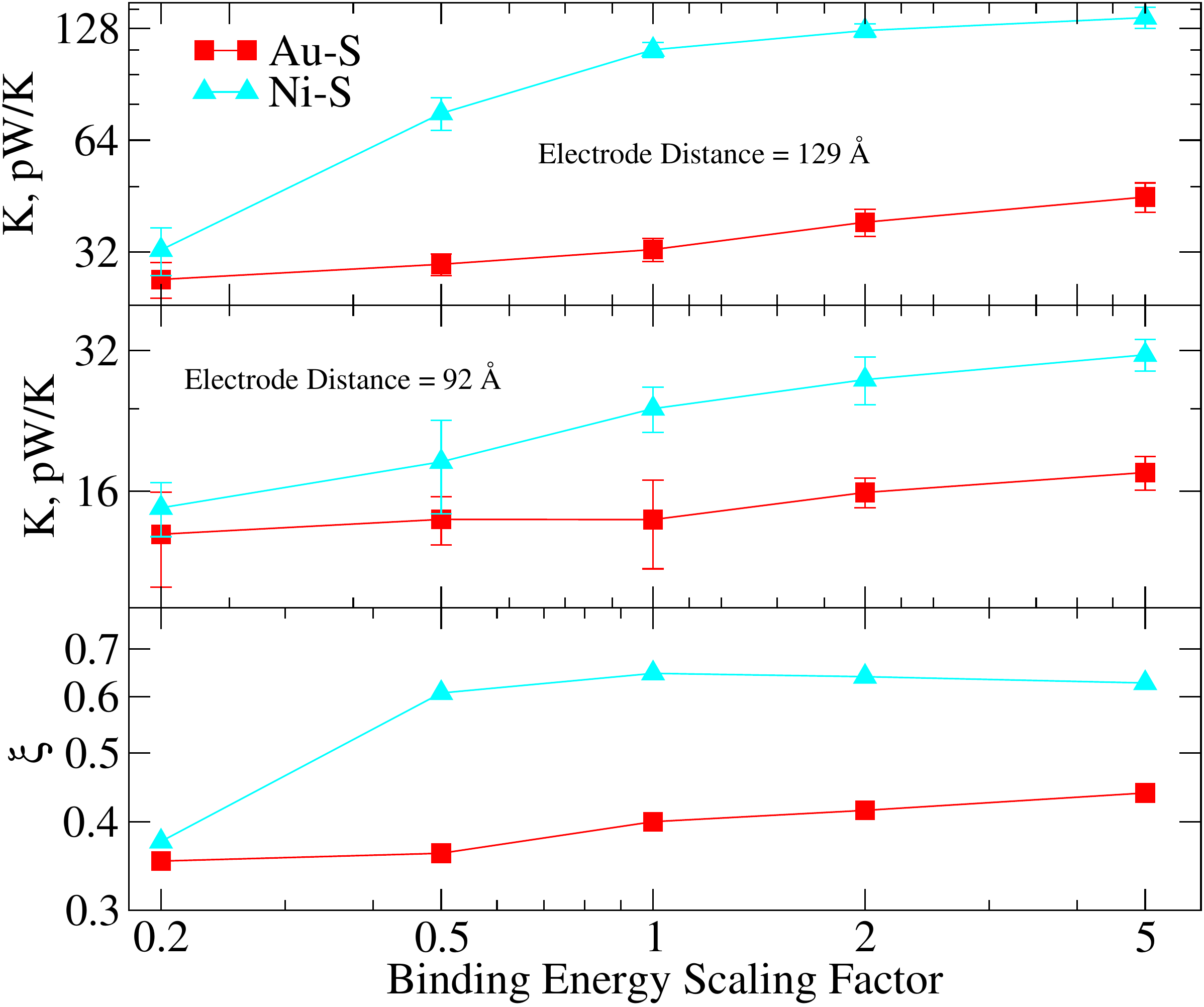}
\caption{The thermal conductance as the metal-molecule coupling varies for Au and Ni metals. The metal-molecule couplings for these metals are varied by scaling the binding energy, $D_e$, in the Morse potential as shown in Table \ref{paramMorse}. The top panel shows the results of the calculations that were carried out at the interelectrode distance of $129.1$ \AA\ while the middle panel shows the results at the interelectrode distance of $91.9$ \AA. Using these calculations, the bottom panel reports the $\xi$ values, $\xi=(K_{\mathrm{H}}-K_{\mathrm{L}})/(K_{\mathrm{H}}+K_{\mathrm{L}})$, where $K_{\mathrm{H}}$ and $K_{\mathrm{L}}$ are the thermal conductances calculated at the interelectrode distances of $129.1$ and $91.9$ \AA, respectively.
}
\label{Kbond}
\end{figure*}

Metal-molecule coupling has significance in the study of the relationship between chemical structure and energy transport properties.\cite{Park2020,Losego2012,Hu2010} 
To study the effect of the metal-polymer bond strength on the thermal energy transport across the interfaces, the binding energy, $D_e$, of the Morse potential, which describes the metal-S bond, was scaled by different values. The
calculated thermal conductance data, obtained from $20$ ns MD simulations,
are plotted in Figure \ref{Kbond}. As can be seen, the thermal conductance values at various electrode distances, in general, increase as the binding energy increases. Note that when the bond strengths increase, the motions of the surface metal atoms and the sulfur atoms are more confined around the potential minimum (equilibrium position), and thus the anharmonic part of the Morse potential contributes less to the average properties of interest including thermal conductance. 
In addition to the bond strength itself, the larger harmonic character of the motion of stronger bonds implies less anharmonic scattering and results in higher thermal conductance.

As can be seen in Fig. \ref{Kbond} the thermal conductance measured for gold leads gradually increases as the binding energy increases. However, the thermal conductance measured for nickel leads initially increases sharply at relatively small values of the binding energy and it almost saturates at strong bonding strengths. In Fig. \ref{Kbond}, comparing the top and middle panels suggests that the effect of binding energy to the metal surfaces is more significant for the stretched polymer chains because there is a relatively larger increase in the thermal conductance for the stretched polymer chains when the binding energy scaling factor increases. Additionally, the bottom panel in Fig. \ref{Kbond} shows that the difference between the thermal conductance of the stretched polymer chains and the compressed ones is less considerable for molecular junctions with relatively weak interactions of the molecule with the metal surface. These observations again suggest that the threshold behavior in the dependence of heat conduction upon stretching is an intrinsic molecular property that reflects phonon scattering event inside the molecular chain, while in the stretched configuration scattering at the metal-substrate interface plays a greater role.

\clearpage
{\it Multi-chain Junctions.}

\begin{figure*}
\includegraphics[width=0.8\linewidth]{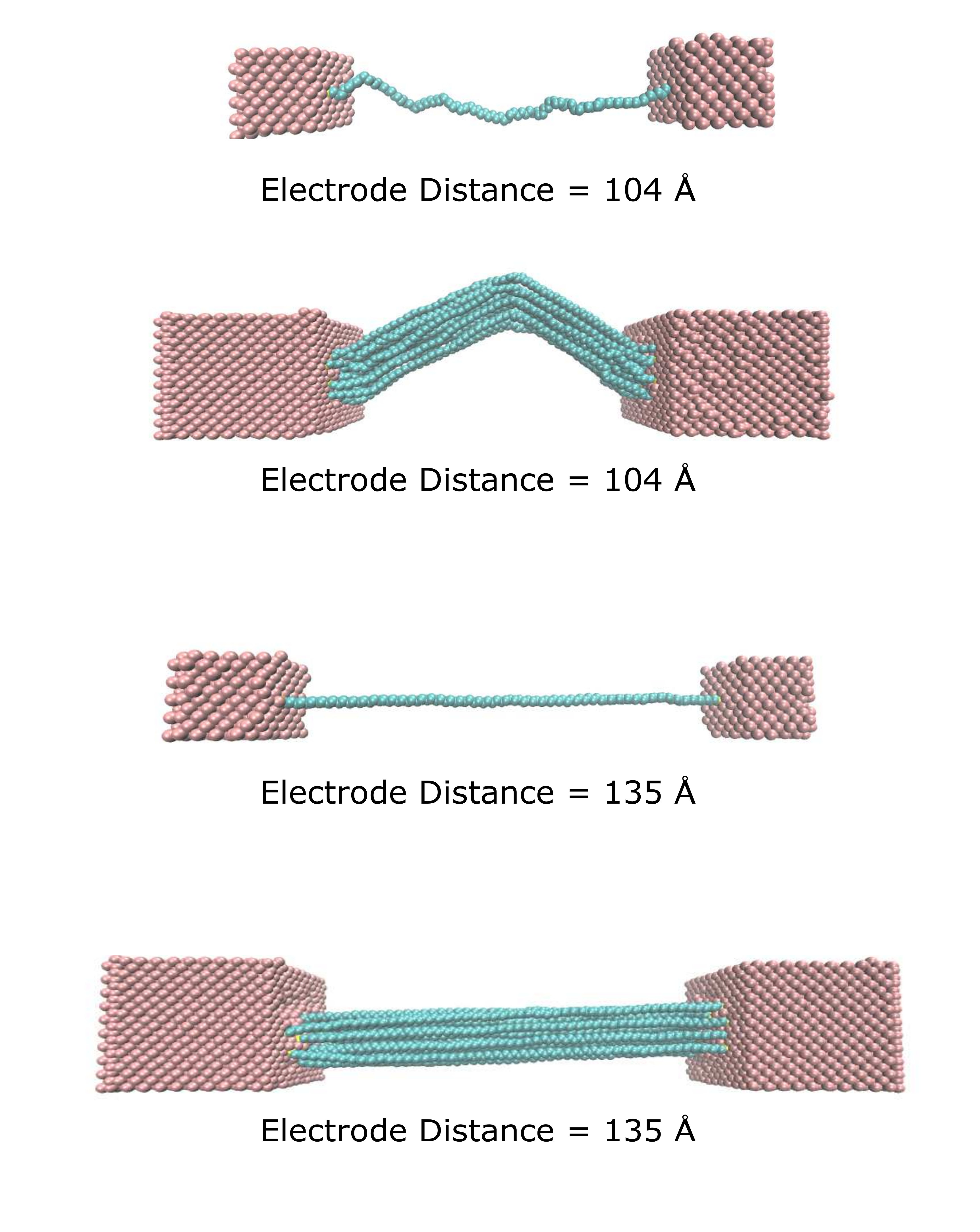}
\caption{Snapshots of single polyethylene (PE) chains and wires consisting of $16$ PE chains sandwiched between gold leads at various electrode distances.}
\label{wireconfig}
\end{figure*}

In this Section, we study the behaviors of molecular wires, comprising several polymer chains under compression and stretching. These wires consist of clusters of parallel polymer chains ($4$ and $16$ polymer chains) sandwiched in between metal leads (see Fig. \ref{wireconfig}).
The initial multi-chain configurations were generated using $2\times2$ and $4\times4$ arrays of the single chain with a sulfur-sulfur spacing of $0.49$ nm.\cite{Ulman1996} However, this distance can change during the MD simulations depending on the force fields used.
Typical distances (between the S sites in two polymers) are found in the range of $4.1$-$5.4$ \AA\ with a maximum around $4.7$ \AA. 
  
The simulation details are similar to those for the single polymer calculations, except that larger metal leads were used for wires consisting of $16$ polymer chains. Note that Lennard-Jones interactions were used for intermolecular interactions between the PE chains in the aforementioned wires and $40$ ns MD simulations were used to report the thermal conductance.  

Figure \ref{wireconfig} shows the snapshots of single PE chains and wires consisting of $16$ PE chains, indicating that stretching/compressing the wires comprising several such chains result in very different configurations than the ones for a single polymer chain. Such differences in configurations are in fact related to the substantial interchain correlations originated from intermolecular interactions. Therefore, the polymer chain nanowires sandwiched between the metal leads are more ordered than the corresponding single polymer chains.
 
\begin{figure*}[tbh]
\includegraphics[width=0.8\linewidth]{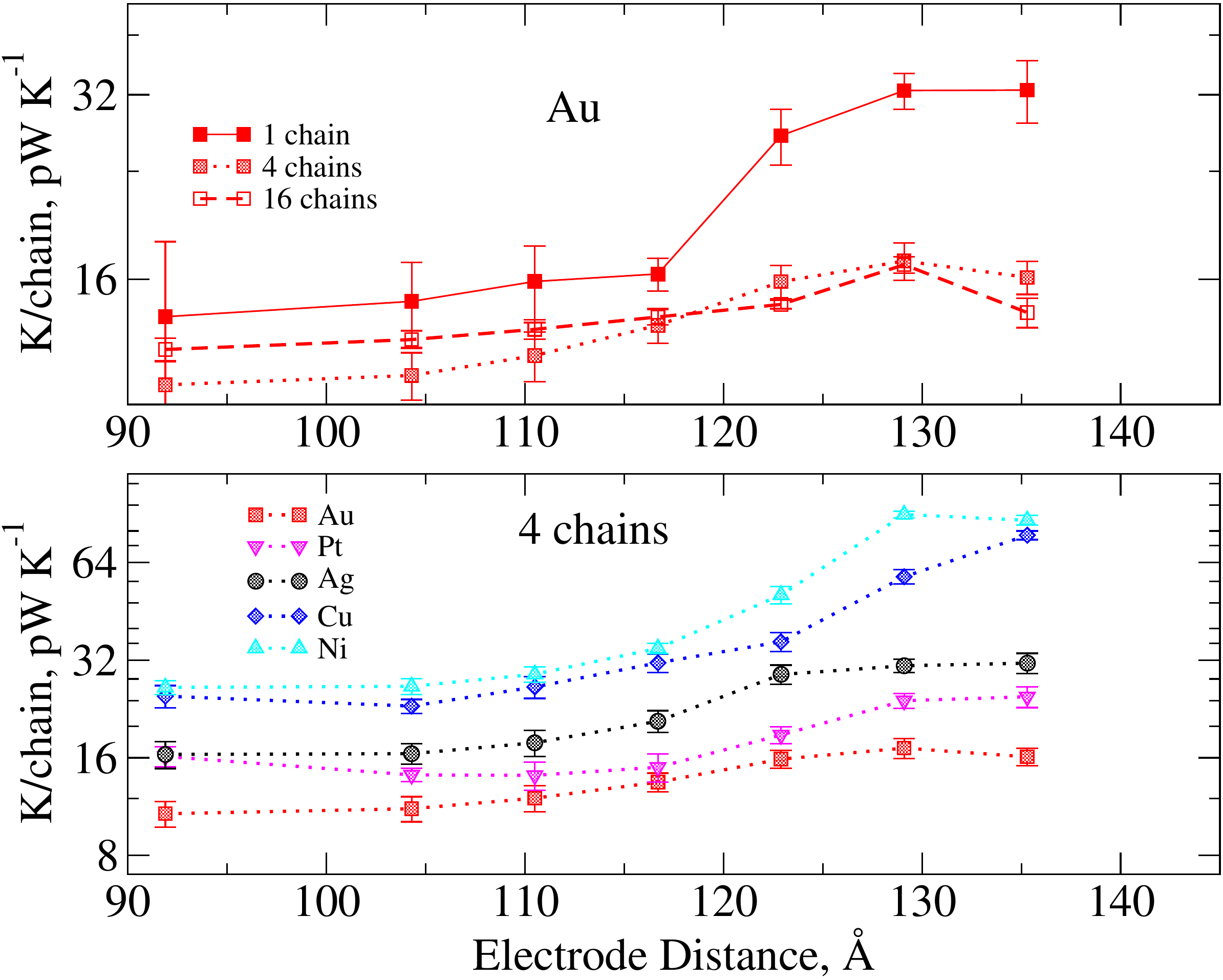}
\caption{Top panel: The thermal conductance per chain for wires of various lengths consisting of $1$, $4$, and $16$ polyethylene (PE) polymer chains when sandwiched in between gold leads. The PE polymer chains consist of $96$ carbon atoms and they are attached to gold substrates by the sulfur atoms. Bottom panel: The thermal conductance per chain at various electrode distances for wires consisting of $4$ polymer chains, where the wires are sandwiched in between different metals at various electrode distances.
}
\label{Kwire}
\end{figure*}

Here, we study the stretching behavior of the heat conduction of wires comprising several chains placed between metal substrates.
The top panel of Fig. \ref{Kwire} compares the thermal conductance per chain for wires consisting of one, four, and sixteen PE polymer chains, all of which are sandwiched in between gold leads. 
In general, the thermal conductance for all wires tends to be constant at the compressed state and increase at critical electrode distances showing the threshold behavior. The magnitude of the thermal conductance for single molecular junctions appear to be greater than the per-chain conductance of the multi-chain junctions. 

The sensitivity of the thermal conductance per chain to the type of metal substrate is investigated for molecular wires consisting of $4$ polymer chains sandwiched between various metals.
The bottom panel of Fig. \ref{Kwire} shows that the effect of substrate is significant in these molecular junctions. Similar to single molecular junctions, the thermal conductance per chain generally increases as the polymer chains are stretched. In addition, the magnitudes of thermal conductances per chain for the Ni multi-chain junctions are higher than other multi-chain junctions although they are smaller than the thermal conductances of the aligned bulk polymers\cite{Dinpajooh2020}. A detailed understanding of such scattering mechanisms as well as possible anharmonic channels for heat transfer are interesting future projects.

\newpage
\pagebreak
\newpage	
\clearpage

\section{Summary \& Conclusions}
\label{conclusion}

We have used classical nonequilibrium molecular dynamics (NEMD) simulations to investigate the room temperature heat conductance behavior of molecular junctions that comprise polyethylene polymer chains connecting various metal leads under different compression/stretching states. We have assumed that the metal electrons couple only with phonons inside the metal\cite{Majumdar2004} so that interfacial heat conductance is dominated by phonon heat transfer. This disregards the possibility that a direct interfacial coupling between the metal electrons and phonons may also exist.\cite{Huberman1994} However, recent studies suggest that at room temperature the electron-phonon coupling is not a major contributor to the thermal resistance across metal-dielectric interfaces.\cite{Singh2013}

We find that as one stretches such simple molecular junctions consisting of linear large alkane chains, the thermal conductance is almost constant at relatively short electrode distances while it tends to increase at critical electrode distances showing a threshold phenomenon (see Fig. \ref{KM}). 
Good correlations between the threshold phenomena and structural parameters such as average CCC angle values and fraction gauche conformations are observed (see Fig. \ref{CCC}). 
However, in a future work it would be interesting to further address the effect of stretching on junctions consisting of very short molecular structures\cite{Cui2019,Klockner2019} and/or more complex molecular structures such as flexible diketone moieties, where conformational switches can occur upon stretching.\cite{Wu2020}

The nature of metal leads has significant effects on the thermal conductance and there is a good correlation between the thermal conductance and the overlap of density of states of the metal leads and the polymer chain (see Table \ref{oval}).
Therefore, the calculated values of
thermal conductance for these molecular junctions at room temperature followed the trend for the density of states overlap, which is in agreement with the existing paradigm based on the harmonic approximation and elastic scattering. However, at higher temperatures we expect that anharmonic effects leading to inelastic phonon transport become more significant and can be a subject of a future study.\cite{Fang2020} 

Importantly, the observed threshold stretching behavior does not depend on the metal-molecule binding and the metal spectral properties, indicating that it is an intrinsic property of heat conduction through the alkane molecular chain. The molecule-metal bond is obviously an important factor controlling the heat conduction properties of such junctions.\cite{Chen2020a,Park2020} Consistent with recent experimental results,\cite{Hopkins2012} we find that stronger bonding character at the interface can indeed lead to higher thermal conductance (see Fig. \ref{Kbond}). 
Consistent with the above conclusion that the threshold stretching behavior is an intrinsic molecular property, we find that this behavior is less pronounced in a junction where heat transport is dominated by the metal-molecule interface. 
Additionally, we compare the thermal conductance of single-molecule junctions with several-molecules junctions. A somewhat similar threshold behavior is observed for several-molecules junctions (see Fig. \ref{Kwire}).

In summary, we have shown that the phonon transport through molecular junctions can be tuned by physical perturbation and a general way to increase the thermal conductance for molecular junctions consisting of simple linear molecular structures is through mechanical stretching. The effect of stretching on the electronic contributions of the heat transfer in molecular junctions remains an open issue.

\section*{Data Availability}
The data that support the findings of this study are available on github at:

\url{https://github.com/dinpajooh/PE_heat}.

\section*{Acknowledgments}
This work has been supported by the U.S. National Science Foundation under Grant No. CHE1953701 and the University of Pennsylvania.
This work used the Extreme Science and Engineering Discovery Environment (XSEDE),\cite{xsede}
which is supported by the National Science Foundation Grant No. ACI-1548562. 
This work used the XSEDE COMET at the San Diego Supercomputer Center through allocation TG-CHE190086.

\clearpage

\bibliographystyle{aipnum4-1}

\begin{thebibliography}{52}%
\makeatletter
\providecommand \@ifxundefined [1]{%
 \@ifx{#1\undefined}
}%
\providecommand \@ifnum [1]{%
 \ifnum #1\expandafter \@firstoftwo
 \else \expandafter \@secondoftwo
 \fi
}%
\providecommand \@ifx [1]{%
 \ifx #1\expandafter \@firstoftwo
 \else \expandafter \@secondoftwo
 \fi
}%
\providecommand \natexlab [1]{#1}%
\providecommand \enquote  [1]{``#1''}%
\providecommand \bibnamefont  [1]{#1}%
\providecommand \bibfnamefont [1]{#1}%
\providecommand \citenamefont [1]{#1}%
\providecommand \href@noop [0]{\@secondoftwo}%
\providecommand \href [0]{\begingroup \@sanitize@url \@href}%
\providecommand \@href[1]{\@@startlink{#1}\@@href}%
\providecommand \@@href[1]{\endgroup#1\@@endlink}%
\providecommand \@sanitize@url [0]{\catcode `\\12\catcode `\$12\catcode
  `\&12\catcode `\#12\catcode `\^12\catcode `\_12\catcode `\%12\relax}%
\providecommand \@@startlink[1]{}%
\providecommand \@@endlink[0]{}%
\providecommand \url  [0]{\begingroup\@sanitize@url \@url }%
\providecommand \@url [1]{\endgroup\@href {#1}{\urlprefix }}%
\providecommand \urlprefix  [0]{URL }%
\providecommand \Eprint [0]{\href }%
\providecommand \doibase [0]{http://dx.doi.org/}%
\providecommand \selectlanguage [0]{\@gobble}%
\providecommand \bibinfo  [0]{\@secondoftwo}%
\providecommand \bibfield  [0]{\@secondoftwo}%
\providecommand \translation [1]{[#1]}%
\providecommand \BibitemOpen [0]{}%
\providecommand \bibitemStop [0]{}%
\providecommand \bibitemNoStop [0]{.\EOS\space}%
\providecommand \EOS [0]{\spacefactor3000\relax}%
\providecommand \BibitemShut  [1]{\csname bibitem#1\endcsname}%
\let\auto@bib@innerbib\@empty
\bibitem [{\citenamefont {Cui}\ \emph {et~al.}(2017)\citenamefont {Cui},
  \citenamefont {Miao}, \citenamefont {Jiang}, \citenamefont {Meyhofer},\ and\
  \citenamefont {Reddy}}]{Cui2017}%
  \BibitemOpen
  \bibfield  {author} {\bibinfo {author} {\bibfnamefont {L.}~\bibnamefont
  {Cui}}, \bibinfo {author} {\bibfnamefont {R.}~\bibnamefont {Miao}}, \bibinfo
  {author} {\bibfnamefont {C.}~\bibnamefont {Jiang}}, \bibinfo {author}
  {\bibfnamefont {E.}~\bibnamefont {Meyhofer}}, \ and\ \bibinfo {author}
  {\bibfnamefont {P.}~\bibnamefont {Reddy}},\ }\href {\doibase
  10.1063/1.4976982} {\bibfield  {journal} {\bibinfo  {journal} {J. Chem.
  Phys.}\ }\textbf {\bibinfo {volume} {146}},\ \bibinfo {pages} {092201}
  (\bibinfo {year} {2017})}\BibitemShut {NoStop}%
\bibitem [{\citenamefont {Wang}, \citenamefont {Meyhofer},\ and\ \citenamefont
  {Reddy}(2020)}]{Wang2020}%
  \BibitemOpen
  \bibfield  {author} {\bibinfo {author} {\bibfnamefont {K.}~\bibnamefont
  {Wang}}, \bibinfo {author} {\bibfnamefont {E.}~\bibnamefont {Meyhofer}}, \
  and\ \bibinfo {author} {\bibfnamefont {P.}~\bibnamefont {Reddy}},\ }\href
  {\doibase 10.1002/adfm.201904534} {\bibfield  {journal} {\bibinfo  {journal}
  {Adv. Funct. Mater}\ }\textbf {\bibinfo {volume} {30}},\ \bibinfo {pages} {1}
  (\bibinfo {year} {2020})}\BibitemShut {NoStop}%
\bibitem [{\citenamefont {Huberman}\ and\ \citenamefont
  {Overhauser}(1994)}]{Huberman1994}%
  \BibitemOpen
  \bibfield  {author} {\bibinfo {author} {\bibfnamefont {M.~L.}\ \bibnamefont
  {Huberman}}\ and\ \bibinfo {author} {\bibfnamefont {A.~W.}\ \bibnamefont
  {Overhauser}},\ }\href {\doibase 10.1103/PhysRevB.50.2865} {\bibfield
  {journal} {\bibinfo  {journal} {Phys. Rev. B}\ }\textbf {\bibinfo {volume}
  {50}},\ \bibinfo {pages} {2865} (\bibinfo {year} {1994})}\BibitemShut
  {NoStop}%
\bibitem [{\citenamefont {Majumdar}\ and\ \citenamefont
  {Reddy}(2004)}]{Majumdar2004}%
  \BibitemOpen
  \bibfield  {author} {\bibinfo {author} {\bibfnamefont {A.}~\bibnamefont
  {Majumdar}}\ and\ \bibinfo {author} {\bibfnamefont {P.}~\bibnamefont
  {Reddy}},\ }\href {\doibase 10.1063/1.1758301} {\bibfield  {journal}
  {\bibinfo  {journal} {Appl. Phys. Lett.}\ }\textbf {\bibinfo {volume} {84}},\
  \bibinfo {pages} {4768} (\bibinfo {year} {2004})}\BibitemShut {NoStop}%
\bibitem [{\citenamefont {Mahan}(2009)}]{Mahan2009}%
  \BibitemOpen
  \bibfield  {author} {\bibinfo {author} {\bibfnamefont {G.~D.}\ \bibnamefont
  {Mahan}},\ }\href {\doibase 10.1103/PhysRevB.79.075408} {\bibfield  {journal}
  {\bibinfo  {journal} {Phys. Rev. B}\ }\textbf {\bibinfo {volume} {79}},\
  \bibinfo {pages} {075408} (\bibinfo {year} {2009})}\BibitemShut {NoStop}%
\bibitem [{\citenamefont {Lombard}, \citenamefont {Detcheverry},\ and\
  \citenamefont {Merabia}(2015)}]{Lombard2015}%
  \BibitemOpen
  \bibfield  {author} {\bibinfo {author} {\bibfnamefont {J.}~\bibnamefont
  {Lombard}}, \bibinfo {author} {\bibfnamefont {F.}~\bibnamefont
  {Detcheverry}}, \ and\ \bibinfo {author} {\bibfnamefont {S.}~\bibnamefont
  {Merabia}},\ }\href {\doibase 10.1088/0953-8984/27/1/015007} {\bibfield
  {journal} {\bibinfo  {journal} {J. Phys.: Condens. Matter}\ }\textbf
  {\bibinfo {volume} {27}},\ \bibinfo {pages} {015007} (\bibinfo {year}
  {2015})},\ \Eprint {http://arxiv.org/abs/1501.03176} {arXiv:1501.03176}
  \BibitemShut {NoStop}%
\bibitem [{\citenamefont {Sadasivam}, \citenamefont {Waghmare},\ and\
  \citenamefont {Fisher}(2015)}]{Sadasivam2015}%
  \BibitemOpen
  \bibfield  {author} {\bibinfo {author} {\bibfnamefont {S.}~\bibnamefont
  {Sadasivam}}, \bibinfo {author} {\bibfnamefont {U.~V.}\ \bibnamefont
  {Waghmare}}, \ and\ \bibinfo {author} {\bibfnamefont {T.~S.}\ \bibnamefont
  {Fisher}},\ }\href {\doibase 10.1063/1.4916729} {\bibfield  {journal}
  {\bibinfo  {journal} {J. Appl. Phys.}\ }\textbf {\bibinfo {volume} {117}},\
  \bibinfo {pages} {134502} (\bibinfo {year} {2015})},\ \Eprint
  {http://arxiv.org/abs/1501.02763} {arXiv:1501.02763} \BibitemShut {NoStop}%
\bibitem [{\citenamefont {Giri}\ and\ \citenamefont
  {Hopkins}(2020)}]{Giri2020}%
  \BibitemOpen
  \bibfield  {author} {\bibinfo {author} {\bibfnamefont {A.}~\bibnamefont
  {Giri}}\ and\ \bibinfo {author} {\bibfnamefont {P.~E.}\ \bibnamefont
  {Hopkins}},\ }\href {\doibase 10.1002/adfm.201903857} {\bibfield  {journal}
  {\bibinfo  {journal} {Adv. Funct. Mater.}\ }\textbf {\bibinfo {volume}
  {30}},\ \bibinfo {pages} {1903857} (\bibinfo {year} {2020})}\BibitemShut
  {NoStop}%
\bibitem [{\citenamefont {Kl{\"{o}}ckner}\ \emph {et~al.}(2016)\citenamefont
  {Kl{\"{o}}ckner}, \citenamefont {B{\"{u}}rkle}, \citenamefont {Cuevas},\ and\
  \citenamefont {Pauly}}]{Klockner2016}%
  \BibitemOpen
  \bibfield  {author} {\bibinfo {author} {\bibfnamefont {J.~C.}\ \bibnamefont
  {Kl{\"{o}}ckner}}, \bibinfo {author} {\bibfnamefont {M.}~\bibnamefont
  {B{\"{u}}rkle}}, \bibinfo {author} {\bibfnamefont {J.~C.}\ \bibnamefont
  {Cuevas}}, \ and\ \bibinfo {author} {\bibfnamefont {F.}~\bibnamefont
  {Pauly}},\ }\href {\doibase 10.1103/PhysRevB.94.205425} {\bibfield  {journal}
  {\bibinfo  {journal} {Phys. Rev. B}\ }\textbf {\bibinfo {volume} {94}},\
  \bibinfo {pages} {205425} (\bibinfo {year} {2016})}\BibitemShut {NoStop}%
\bibitem [{\citenamefont {Li}\ \emph {et~al.}(2015)\citenamefont {Li},
  \citenamefont {Duchemin}, \citenamefont {Xiong}, \citenamefont {Solomon},\
  and\ \citenamefont {Donadio}}]{Li2015}%
  \BibitemOpen
  \bibfield  {author} {\bibinfo {author} {\bibfnamefont {Q.}~\bibnamefont
  {Li}}, \bibinfo {author} {\bibfnamefont {I.}~\bibnamefont {Duchemin}},
  \bibinfo {author} {\bibfnamefont {S.}~\bibnamefont {Xiong}}, \bibinfo
  {author} {\bibfnamefont {G.~C.}\ \bibnamefont {Solomon}}, \ and\ \bibinfo
  {author} {\bibfnamefont {D.}~\bibnamefont {Donadio}},\ }\href {\doibase
  10.1021/acs.jpcc.5b07429} {\bibfield  {journal} {\bibinfo  {journal} {J.
  Phys. Chem. C}\ }\textbf {\bibinfo {volume} {119}},\ \bibinfo {pages} {24636}
  (\bibinfo {year} {2015})},\ \Eprint {http://arxiv.org/abs/1511.06058}
  {arXiv:1511.06058} \BibitemShut {NoStop}%
\bibitem [{\citenamefont {Vacek}\ \emph {et~al.}(2015)\citenamefont {Vacek},
  \citenamefont {Chocholou{\v{s}}ov{\'{a}}}, \citenamefont {Star{\'{a}}},
  \citenamefont {Star{\'{y}}},\ and\ \citenamefont {Dubi}}]{Vacek2015}%
  \BibitemOpen
  \bibfield  {author} {\bibinfo {author} {\bibfnamefont {J.}~\bibnamefont
  {Vacek}}, \bibinfo {author} {\bibfnamefont {J.~V.}\ \bibnamefont
  {Chocholou{\v{s}}ov{\'{a}}}}, \bibinfo {author} {\bibfnamefont {I.~G.}\
  \bibnamefont {Star{\'{a}}}}, \bibinfo {author} {\bibfnamefont
  {I.}~\bibnamefont {Star{\'{y}}}}, \ and\ \bibinfo {author} {\bibfnamefont
  {Y.}~\bibnamefont {Dubi}},\ }\href {\doibase 10.1039/c5nr01297j} {\bibfield
  {journal} {\bibinfo  {journal} {Nanoscale}\ }\textbf {\bibinfo {volume}
  {7}},\ \bibinfo {pages} {8793} (\bibinfo {year} {2015})}\BibitemShut
  {NoStop}%
\bibitem [{\citenamefont {Duli{\'{c}}}\ \emph {et~al.}(2003)\citenamefont
  {Duli{\'{c}}}, \citenamefont {van~der Molen}, \citenamefont {Kudernac},
  \citenamefont {Jonkman}, \citenamefont {de~Jong}, \citenamefont {Bowden},
  \citenamefont {van Esch}, \citenamefont {Feringa},\ and\ \citenamefont {van
  Wees}}]{Dulic2003}%
  \BibitemOpen
  \bibfield  {author} {\bibinfo {author} {\bibfnamefont {D.}~\bibnamefont
  {Duli{\'{c}}}}, \bibinfo {author} {\bibfnamefont {S.~J.}\ \bibnamefont
  {van~der Molen}}, \bibinfo {author} {\bibfnamefont {T.}~\bibnamefont
  {Kudernac}}, \bibinfo {author} {\bibfnamefont {H.~T.}\ \bibnamefont
  {Jonkman}}, \bibinfo {author} {\bibfnamefont {J.~J.}\ \bibnamefont
  {de~Jong}}, \bibinfo {author} {\bibfnamefont {T.~N.}\ \bibnamefont {Bowden}},
  \bibinfo {author} {\bibfnamefont {J.}~\bibnamefont {van Esch}}, \bibinfo
  {author} {\bibfnamefont {B.~L.}\ \bibnamefont {Feringa}}, \ and\ \bibinfo
  {author} {\bibfnamefont {B.~J.}\ \bibnamefont {van Wees}},\ }\href {\doibase
  10.1103/PhysRevLett.91.207402} {\bibfield  {journal} {\bibinfo  {journal}
  {Phys. Rev. Lett.}\ }\textbf {\bibinfo {volume} {91}},\ \bibinfo {pages}
  {207402} (\bibinfo {year} {2003})}\BibitemShut {NoStop}%
\bibitem [{\citenamefont {Cui}\ \emph {et~al.}(2019)\citenamefont {Cui},
  \citenamefont {Hur}, \citenamefont {Akbar}, \citenamefont {Kl{\"{o}}ckner},
  \citenamefont {Jeong}, \citenamefont {Pauly}, \citenamefont {Jang},
  \citenamefont {Reddy},\ and\ \citenamefont {Meyhofer}}]{Cui2019}%
  \BibitemOpen
  \bibfield  {author} {\bibinfo {author} {\bibfnamefont {L.}~\bibnamefont
  {Cui}}, \bibinfo {author} {\bibfnamefont {S.}~\bibnamefont {Hur}}, \bibinfo
  {author} {\bibfnamefont {Z.~A.}\ \bibnamefont {Akbar}}, \bibinfo {author}
  {\bibfnamefont {J.~C.}\ \bibnamefont {Kl{\"{o}}ckner}}, \bibinfo {author}
  {\bibfnamefont {W.}~\bibnamefont {Jeong}}, \bibinfo {author} {\bibfnamefont
  {F.}~\bibnamefont {Pauly}}, \bibinfo {author} {\bibfnamefont {S.-Y.}\
  \bibnamefont {Jang}}, \bibinfo {author} {\bibfnamefont {P.}~\bibnamefont
  {Reddy}}, \ and\ \bibinfo {author} {\bibfnamefont {E.}~\bibnamefont
  {Meyhofer}},\ }\href {\doibase 10.1038/s41586-019-1420-z} {\bibfield
  {journal} {\bibinfo  {journal} {Nature}\ }\textbf {\bibinfo {volume} {572}},\
  \bibinfo {pages} {628} (\bibinfo {year} {2019})}\BibitemShut {NoStop}%
\bibitem [{\citenamefont {Liu}\ and\ \citenamefont {Yang}(2010)}]{Liu2010}%
  \BibitemOpen
  \bibfield  {author} {\bibinfo {author} {\bibfnamefont {J.}~\bibnamefont
  {Liu}}\ and\ \bibinfo {author} {\bibfnamefont {R.}~\bibnamefont {Yang}},\
  }\href {\doibase 10.1103/PhysRevB.81.174122} {\bibfield  {journal} {\bibinfo
  {journal} {Phys. Rev. B}\ }\textbf {\bibinfo {volume} {81}},\ \bibinfo
  {pages} {174122} (\bibinfo {year} {2010})}\BibitemShut {NoStop}%
\bibitem [{\citenamefont {Liu}\ and\ \citenamefont {Yang}(2012)}]{Liu2012}%
  \BibitemOpen
  \bibfield  {author} {\bibinfo {author} {\bibfnamefont {J.}~\bibnamefont
  {Liu}}\ and\ \bibinfo {author} {\bibfnamefont {R.}~\bibnamefont {Yang}},\
  }\href {\doibase 10.1103/PhysRevB.86.104307} {\bibfield  {journal} {\bibinfo
  {journal} {Phys. Rev. B}\ }\textbf {\bibinfo {volume} {86}},\ \bibinfo
  {pages} {104307} (\bibinfo {year} {2012})}\BibitemShut {NoStop}%
\bibitem [{\citenamefont {Chen}\ and\ \citenamefont {Chen}(2020)}]{Chen2020a}%
  \BibitemOpen
  \bibfield  {author} {\bibinfo {author} {\bibfnamefont {X.~K.}\ \bibnamefont
  {Chen}}\ and\ \bibinfo {author} {\bibfnamefont {K.~Q.}\ \bibnamefont
  {Chen}},\ }\href {\doibase 10.1088/1361-648X/ab5e57} {\bibfield  {journal}
  {\bibinfo  {journal} {J. Phys.: Condens Matt.}\ }\textbf {\bibinfo {volume}
  {32}},\ \bibinfo {pages} {153002} (\bibinfo {year} {2020})}\BibitemShut
  {NoStop}%
\bibitem [{\citenamefont {Dinpajooh}\ and\ \citenamefont
  {Nitzan}(2020)}]{Dinpajooh2020}%
  \BibitemOpen
  \bibfield  {author} {\bibinfo {author} {\bibfnamefont {M.}~\bibnamefont
  {Dinpajooh}}\ and\ \bibinfo {author} {\bibfnamefont {A.}~\bibnamefont
  {Nitzan}},\ }\href {\doibase 10.1063/5.0023085} {\bibfield  {journal}
  {\bibinfo  {journal} {J. Chem. Phys.}\ }\textbf {\bibinfo {volume} {153}}
  (\bibinfo {year} {2020}),\ 10.1063/5.0023085}\BibitemShut {NoStop}%
\bibitem [{\citenamefont {Zhang}\ and\ \citenamefont {Luo}(2012)}]{Zhang2012}%
  \BibitemOpen
  \bibfield  {author} {\bibinfo {author} {\bibfnamefont {T.}~\bibnamefont
  {Zhang}}\ and\ \bibinfo {author} {\bibfnamefont {T.}~\bibnamefont {Luo}},\
  }\href {\doibase 10.1063/1.4759293} {\bibfield  {journal} {\bibinfo
  {journal} {J. Appl. Phys.}\ }\textbf {\bibinfo {volume} {112}},\ \bibinfo
  {pages} {094304} (\bibinfo {year} {2012})}\BibitemShut {NoStop}%
\bibitem [{\citenamefont {Mosso}\ \emph {et~al.}(2019)\citenamefont {Mosso},
  \citenamefont {Sadeghi}, \citenamefont {Gemma}, \citenamefont {Sangtarash},
  \citenamefont {Drechsler}, \citenamefont {Lambert},\ and\ \citenamefont
  {Gotsmann}}]{Mosso2019}%
  \BibitemOpen
  \bibfield  {author} {\bibinfo {author} {\bibfnamefont {N.}~\bibnamefont
  {Mosso}}, \bibinfo {author} {\bibfnamefont {H.}~\bibnamefont {Sadeghi}},
  \bibinfo {author} {\bibfnamefont {A.}~\bibnamefont {Gemma}}, \bibinfo
  {author} {\bibfnamefont {S.}~\bibnamefont {Sangtarash}}, \bibinfo {author}
  {\bibfnamefont {U.}~\bibnamefont {Drechsler}}, \bibinfo {author}
  {\bibfnamefont {C.}~\bibnamefont {Lambert}}, \ and\ \bibinfo {author}
  {\bibfnamefont {B.}~\bibnamefont {Gotsmann}},\ }\href {\doibase
  10.1021/acs.nanolett.9b02089} {\bibfield  {journal} {\bibinfo  {journal}
  {Nano Lett.}\ }\textbf {\bibinfo {volume} {19}},\ \bibinfo {pages} {7614}
  (\bibinfo {year} {2019})}\BibitemShut {NoStop}%
\bibitem [{\citenamefont {Landry}\ and\ \citenamefont
  {McGaughey}(2009)}]{Landry2009}%
  \BibitemOpen
  \bibfield  {author} {\bibinfo {author} {\bibfnamefont {E.~S.}\ \bibnamefont
  {Landry}}\ and\ \bibinfo {author} {\bibfnamefont {A.~J.~H.}\ \bibnamefont
  {McGaughey}},\ }\href {\doibase 10.1103/PhysRevB.80.165304} {\bibfield
  {journal} {\bibinfo  {journal} {Phys. Rev. B}\ }\textbf {\bibinfo {volume}
  {80}},\ \bibinfo {pages} {165304} (\bibinfo {year} {2009})}\BibitemShut
  {NoStop}%
\bibitem [{\citenamefont {{Moghaddasi Fereidani}}\ and\ \citenamefont
  {Segal}(2019)}]{MoghaddasiFereidani2019}%
  \BibitemOpen
  \bibfield  {author} {\bibinfo {author} {\bibfnamefont {R.}~\bibnamefont
  {{Moghaddasi Fereidani}}}\ and\ \bibinfo {author} {\bibfnamefont
  {D.}~\bibnamefont {Segal}},\ }\href {\doibase 10.1063/1.5075620} {\bibfield
  {journal} {\bibinfo  {journal} {J. Chem. Phys.}\ }\textbf {\bibinfo {volume}
  {150}},\ \bibinfo {pages} {024105} (\bibinfo {year} {2019})},\ \Eprint
  {http://arxiv.org/abs/1810.09888} {arXiv:1810.09888} \BibitemShut {NoStop}%
\bibitem [{\citenamefont {Plimpton}(1995)}]{Plimpton1995}%
  \BibitemOpen
  \bibfield  {author} {\bibinfo {author} {\bibfnamefont {S.}~\bibnamefont
  {Plimpton}},\ }\href@noop {} {\bibfield  {journal} {\bibinfo  {journal} {J.
  Comp. Phys.}\ }\textbf {\bibinfo {volume} {117}},\ \bibinfo {pages} {1}
  (\bibinfo {year} {1995})}\BibitemShut {NoStop}%
\bibitem [{\citenamefont {Boone}, \citenamefont {Babaei},\ and\ \citenamefont
  {Wilmer}(2019)}]{Boone2019}%
  \BibitemOpen
  \bibfield  {author} {\bibinfo {author} {\bibfnamefont {P.}~\bibnamefont
  {Boone}}, \bibinfo {author} {\bibfnamefont {H.}~\bibnamefont {Babaei}}, \
  and\ \bibinfo {author} {\bibfnamefont {C.~E.}\ \bibnamefont {Wilmer}},\
  }\href {\doibase 10.1021/acs.jctc.9b00252} {\bibfield  {journal} {\bibinfo
  {journal} {J. Chem. Theory Comput.}\ }\textbf {\bibinfo {volume} {15}},\
  \bibinfo {pages} {5579} (\bibinfo {year} {2019})}\BibitemShut {NoStop}%
\bibitem [{\citenamefont {Daw}\ and\ \citenamefont {Baskes}(1984)}]{Daw1984}%
  \BibitemOpen
  \bibfield  {author} {\bibinfo {author} {\bibfnamefont {M.~S.}\ \bibnamefont
  {Daw}}\ and\ \bibinfo {author} {\bibfnamefont {M.~I.}\ \bibnamefont
  {Baskes}},\ }\href {\doibase 10.1103/PhysRevB.29.6443} {\bibfield  {journal}
  {\bibinfo  {journal} {Phys. Rev. B}\ }\textbf {\bibinfo {volume} {29}},\
  \bibinfo {pages} {6443} (\bibinfo {year} {1984})}\BibitemShut {NoStop}%
\bibitem [{\citenamefont {Foiles}, \citenamefont {Baskes},\ and\ \citenamefont
  {Daw}(1986)}]{Foiles1986}%
  \BibitemOpen
  \bibfield  {author} {\bibinfo {author} {\bibfnamefont {S.~M.}\ \bibnamefont
  {Foiles}}, \bibinfo {author} {\bibfnamefont {M.~I.}\ \bibnamefont {Baskes}},
  \ and\ \bibinfo {author} {\bibfnamefont {M.~S.}\ \bibnamefont {Daw}},\ }\href
  {\doibase 10.1103/PhysRevB.33.7983} {\bibfield  {journal} {\bibinfo
  {journal} {Phys. Rev. B}\ }\textbf {\bibinfo {volume} {33}},\ \bibinfo
  {pages} {7983} (\bibinfo {year} {1986})}\BibitemShut {NoStop}%
\bibitem [{\citenamefont {Ackland}, \citenamefont {Finnis},\ and\ \citenamefont
  {Vitek}(1988)}]{Ackland1988}%
  \BibitemOpen
  \bibfield  {author} {\bibinfo {author} {\bibfnamefont {G.~J.}\ \bibnamefont
  {Ackland}}, \bibinfo {author} {\bibfnamefont {M.~W.}\ \bibnamefont {Finnis}},
  \ and\ \bibinfo {author} {\bibfnamefont {V.}~\bibnamefont {Vitek}},\ }\href
  {\doibase 10.1088/0305-4608/18/8/002} {\bibfield  {journal} {\bibinfo
  {journal} {J. Phys. F.: Met. Phys.}\ }\textbf {\bibinfo {volume} {18}},\
  \bibinfo {pages} {L153} (\bibinfo {year} {1988})}\BibitemShut {NoStop}%
\bibitem [{\citenamefont {Martin}\ and\ \citenamefont
  {Siepmann}(1998)}]{Martin1998}%
  \BibitemOpen
  \bibfield  {author} {\bibinfo {author} {\bibfnamefont {M.~G.}\ \bibnamefont
  {Martin}}\ and\ \bibinfo {author} {\bibfnamefont {J.~I.}\ \bibnamefont
  {Siepmann}},\ }\href {\doibase 10.1021/jp972543+} {\bibfield  {journal}
  {\bibinfo  {journal} {J. Phys. Chem. B}\ }\textbf {\bibinfo {volume} {102}},\
  \bibinfo {pages} {2569} (\bibinfo {year} {1998})}\BibitemShut {NoStop}%
\bibitem [{\citenamefont {Wick}, \citenamefont {Martin},\ and\ \citenamefont
  {Siepmann}(2000)}]{Wick2000}%
  \BibitemOpen
  \bibfield  {author} {\bibinfo {author} {\bibfnamefont {C.~D.}\ \bibnamefont
  {Wick}}, \bibinfo {author} {\bibfnamefont {M.~G.}\ \bibnamefont {Martin}}, \
  and\ \bibinfo {author} {\bibfnamefont {J.~I.}\ \bibnamefont {Siepmann}},\
  }\href {\doibase 10.1021/jp001044x} {\bibfield  {journal} {\bibinfo
  {journal} {J. Phys. Chem. B}\ }\textbf {\bibinfo {volume} {104}},\ \bibinfo
  {pages} {8008} (\bibinfo {year} {2000})}\BibitemShut {NoStop}%
\bibitem [{\citenamefont {Jiang}(2002)}]{Jiang2002}%
  \BibitemOpen
  \bibfield  {author} {\bibinfo {author} {\bibfnamefont {S.}~\bibnamefont
  {Jiang}},\ }\href {\doibase 10.1080/00268970210130948} {\bibfield  {journal}
  {\bibinfo  {journal} {Mol. Phys.}\ }\textbf {\bibinfo {volume} {100}},\
  \bibinfo {pages} {2261} (\bibinfo {year} {2002})}\BibitemShut {NoStop}%
\bibitem [{\citenamefont {Liu}\ \emph {et~al.}(1999)\citenamefont {Liu},
  \citenamefont {Yong}, \citenamefont {Garrison},\ and\ \citenamefont
  {Vickerman}}]{Liu1999}%
  \BibitemOpen
  \bibfield  {author} {\bibinfo {author} {\bibfnamefont {K.~S.}\ \bibnamefont
  {Liu}}, \bibinfo {author} {\bibfnamefont {C.~W.}\ \bibnamefont {Yong}},
  \bibinfo {author} {\bibfnamefont {B.~J.}\ \bibnamefont {Garrison}}, \ and\
  \bibinfo {author} {\bibfnamefont {J.~C.}\ \bibnamefont {Vickerman}},\ }\href
  {\doibase 10.1021/jp984071k} {\bibfield  {journal} {\bibinfo  {journal} {J.
  Phys. Chem. B}\ }\textbf {\bibinfo {volume} {103}},\ \bibinfo {pages} {3195}
  (\bibinfo {year} {1999})}\BibitemShut {NoStop}%
\bibitem [{\citenamefont {Yang}, \citenamefont {Sun},\ and\ \citenamefont
  {Deng}(2019)}]{Yang2019}%
  \BibitemOpen
  \bibfield  {author} {\bibinfo {author} {\bibfnamefont {L.}~\bibnamefont
  {Yang}}, \bibinfo {author} {\bibfnamefont {L.}~\bibnamefont {Sun}}, \ and\
  \bibinfo {author} {\bibfnamefont {W.~Q.}\ \bibnamefont {Deng}},\ }\href
  {\doibase 10.1021/acs.jpca.9b02055} {\bibfield  {journal} {\bibinfo
  {journal} {J. Phys. Chem. A}\ }\textbf {\bibinfo {volume} {123}},\ \bibinfo
  {pages} {7847} (\bibinfo {year} {2019})}\BibitemShut {NoStop}%
\bibitem [{\citenamefont {Pamuk}\ and\ \citenamefont
  {HalicioÇ§lu}(1976)}]{Pamuk1976}%
  \BibitemOpen
  \bibfield  {author} {\bibinfo {author} {\bibfnamefont {H.}~\bibnamefont
  {Pamuk}}\ and\ \bibinfo {author} {\bibfnamefont {T.}~\bibnamefont
  {HalicioÇ§lu}},\ }\href {\doibase 10.1002/pssa.2210370242} {\bibfield
  {journal} {\bibinfo  {journal} {Phys. Stat. Sol. (a)}\ }\textbf {\bibinfo
  {volume} {37}},\ \bibinfo {pages} {695} (\bibinfo {year} {1976})}\BibitemShut
  {NoStop}%
\bibitem [{\citenamefont {Rapp{\'{e}}}\ \emph {et~al.}(1992)\citenamefont
  {Rapp{\'{e}}}, \citenamefont {Casewit}, \citenamefont {Colwell},
  \citenamefont {Goddard},\ and\ \citenamefont {Skiff}}]{Rappe1992}%
  \BibitemOpen
  \bibfield  {author} {\bibinfo {author} {\bibfnamefont {A.~K.}\ \bibnamefont
  {Rapp{\'{e}}}}, \bibinfo {author} {\bibfnamefont {C.~J.}\ \bibnamefont
  {Casewit}}, \bibinfo {author} {\bibfnamefont {K.~S.}\ \bibnamefont
  {Colwell}}, \bibinfo {author} {\bibfnamefont {W.~A.}\ \bibnamefont
  {Goddard}}, \ and\ \bibinfo {author} {\bibfnamefont {W.~M.}\ \bibnamefont
  {Skiff}},\ }\href {\doibase 10.1021/ja00051a040} {\bibfield  {journal}
  {\bibinfo  {journal} {J. Am. Chem. Soc.}\ }\textbf {\bibinfo {volume}
  {114}},\ \bibinfo {pages} {10024} (\bibinfo {year} {1992})}\BibitemShut
  {NoStop}%
\bibitem [{\citenamefont {Hopkins}, \citenamefont {Duda},\ and\ \citenamefont
  {Norris}(2011)}]{Hopkins2011}%
  \BibitemOpen
  \bibfield  {author} {\bibinfo {author} {\bibfnamefont {P.~E.}\ \bibnamefont
  {Hopkins}}, \bibinfo {author} {\bibfnamefont {J.~C.}\ \bibnamefont {Duda}}, \
  and\ \bibinfo {author} {\bibfnamefont {P.~M.}\ \bibnamefont {Norris}},\
  }\href {\doibase 10.1115/1.4003549} {\bibfield  {journal} {\bibinfo
  {journal} {J. Heat Transfer}\ }\textbf {\bibinfo {volume} {133}},\ \bibinfo
  {pages} {062401} (\bibinfo {year} {2011})}\BibitemShut {NoStop}%
\bibitem [{\citenamefont {S{\"{a}}{\"{a}}skilahti}\ \emph
  {et~al.}(2014)\citenamefont {S{\"{a}}{\"{a}}skilahti}, \citenamefont
  {Oksanen}, \citenamefont {Tulkki},\ and\ \citenamefont
  {Volz}}]{Saaskilahti2014}%
  \BibitemOpen
  \bibfield  {author} {\bibinfo {author} {\bibfnamefont {K.}~\bibnamefont
  {S{\"{a}}{\"{a}}skilahti}}, \bibinfo {author} {\bibfnamefont
  {J.}~\bibnamefont {Oksanen}}, \bibinfo {author} {\bibfnamefont
  {J.}~\bibnamefont {Tulkki}}, \ and\ \bibinfo {author} {\bibfnamefont
  {S.}~\bibnamefont {Volz}},\ }\href {\doibase 10.1103/PhysRevB.90.134312}
  {\bibfield  {journal} {\bibinfo  {journal} {Phys. Rev. B}\ }\textbf {\bibinfo
  {volume} {90}},\ \bibinfo {pages} {134312} (\bibinfo {year} {2014})},\
  \Eprint {http://arxiv.org/abs/1405.3868} {arXiv:1405.3868} \BibitemShut
  {NoStop}%
\bibitem [{\citenamefont {Shen}\ \emph {et~al.}(2011)\citenamefont {Shen},
  \citenamefont {Evans}, \citenamefont {Cahill},\ and\ \citenamefont
  {Keblinski}}]{Shen2011}%
  \BibitemOpen
  \bibfield  {author} {\bibinfo {author} {\bibfnamefont {M.}~\bibnamefont
  {Shen}}, \bibinfo {author} {\bibfnamefont {W.~J.}\ \bibnamefont {Evans}},
  \bibinfo {author} {\bibfnamefont {D.}~\bibnamefont {Cahill}}, \ and\ \bibinfo
  {author} {\bibfnamefont {P.}~\bibnamefont {Keblinski}},\ }\href {\doibase
  10.1103/PhysRevB.84.195432} {\bibfield  {journal} {\bibinfo  {journal} {Phys.
  Rev. B}\ }\textbf {\bibinfo {volume} {84}},\ \bibinfo {pages} {195432}
  (\bibinfo {year} {2011})}\BibitemShut {NoStop}%
\bibitem [{\citenamefont {Lu}, \citenamefont {Chaka},\ and\ \citenamefont
  {Sushko}(2020)}]{Lu2020}%
  \BibitemOpen
  \bibfield  {author} {\bibinfo {author} {\bibfnamefont {Z.}~\bibnamefont
  {Lu}}, \bibinfo {author} {\bibfnamefont {A.~M.}\ \bibnamefont {Chaka}}, \
  and\ \bibinfo {author} {\bibfnamefont {P.~V.}\ \bibnamefont {Sushko}},\
  }\href {\doibase 10.1103/PhysRevB.102.075449} {\bibfield  {journal} {\bibinfo
   {journal} {Phys. Rev. B}\ }\textbf {\bibinfo {volume} {102}},\ \bibinfo
  {pages} {075449} (\bibinfo {year} {2020})}\BibitemShut {NoStop}%
\bibitem [{\citenamefont {O'Brien}\ \emph {et~al.}(2013)\citenamefont
  {O'Brien}, \citenamefont {Shenogin}, \citenamefont {Liu}, \citenamefont
  {Chow}, \citenamefont {Laurencin}, \citenamefont {Mutin}, \citenamefont
  {Yamaguchi}, \citenamefont {Keblinski},\ and\ \citenamefont
  {Ramanath}}]{OBrien2013}%
  \BibitemOpen
  \bibfield  {author} {\bibinfo {author} {\bibfnamefont {P.~J.}\ \bibnamefont
  {O'Brien}}, \bibinfo {author} {\bibfnamefont {S.}~\bibnamefont {Shenogin}},
  \bibinfo {author} {\bibfnamefont {J.}~\bibnamefont {Liu}}, \bibinfo {author}
  {\bibfnamefont {P.~K.}\ \bibnamefont {Chow}}, \bibinfo {author}
  {\bibfnamefont {D.}~\bibnamefont {Laurencin}}, \bibinfo {author}
  {\bibfnamefont {P.~H.}\ \bibnamefont {Mutin}}, \bibinfo {author}
  {\bibfnamefont {M.}~\bibnamefont {Yamaguchi}}, \bibinfo {author}
  {\bibfnamefont {P.}~\bibnamefont {Keblinski}}, \ and\ \bibinfo {author}
  {\bibfnamefont {G.}~\bibnamefont {Ramanath}},\ }\href {\doibase
  10.1038/nmat3465} {\bibfield  {journal} {\bibinfo  {journal} {Nat. Mater.}\
  }\textbf {\bibinfo {volume} {12}},\ \bibinfo {pages} {118} (\bibinfo {year}
  {2013})}\BibitemShut {NoStop}%
\bibitem [{\citenamefont {Park}\ \emph {et~al.}(2020)\citenamefont {Park},
  \citenamefont {Jang}, \citenamefont {Kim}, \citenamefont {Park},
  \citenamefont {Kim},\ and\ \citenamefont {Yoon}}]{Park2020}%
  \BibitemOpen
  \bibfield  {author} {\bibinfo {author} {\bibfnamefont {S.}~\bibnamefont
  {Park}}, \bibinfo {author} {\bibfnamefont {J.}~\bibnamefont {Jang}}, \bibinfo
  {author} {\bibfnamefont {H.}~\bibnamefont {Kim}}, \bibinfo {author}
  {\bibfnamefont {D.~I.}\ \bibnamefont {Park}}, \bibinfo {author}
  {\bibfnamefont {K.}~\bibnamefont {Kim}}, \ and\ \bibinfo {author}
  {\bibfnamefont {H.~J.}\ \bibnamefont {Yoon}},\ }\href {\doibase
  10.1039/d0ta07095e} {\bibfield  {journal} {\bibinfo  {journal} {J. Mater.
  Chem. A}\ }\textbf {\bibinfo {volume} {8}},\ \bibinfo {pages} {19746}
  (\bibinfo {year} {2020})}\BibitemShut {NoStop}%
\bibitem [{\citenamefont {Luo}\ and\ \citenamefont {Lloyd}(2010)}]{Luo2010}%
  \BibitemOpen
  \bibfield  {author} {\bibinfo {author} {\bibfnamefont {T.}~\bibnamefont
  {Luo}}\ and\ \bibinfo {author} {\bibfnamefont {J.~R.}\ \bibnamefont
  {Lloyd}},\ }\href {\doibase 10.1016/j.ijheatmasstransfer.2009.10.033}
  {\bibfield  {journal} {\bibinfo  {journal} {Int. J. Heat Mass Transf.}\
  }\textbf {\bibinfo {volume} {53}},\ \bibinfo {pages} {1} (\bibinfo {year}
  {2010})}\BibitemShut {NoStop}%
\bibitem [{\citenamefont {Luo}\ and\ \citenamefont {Lloyd}(2011)}]{Luo2011a}%
  \BibitemOpen
  \bibfield  {author} {\bibinfo {author} {\bibfnamefont {T.}~\bibnamefont
  {Luo}}\ and\ \bibinfo {author} {\bibfnamefont {J.~R.}\ \bibnamefont
  {Lloyd}},\ }\href {\doibase 10.1063/1.3530685} {\bibfield  {journal}
  {\bibinfo  {journal} {J. Appl. Phys.}\ }\textbf {\bibinfo {volume} {109}},\
  \bibinfo {pages} {034301} (\bibinfo {year} {2011})}\BibitemShut {NoStop}%
\bibitem [{\citenamefont {Reid}, \citenamefont {Pandey},\ and\ \citenamefont
  {Leitner}(2019)}]{Reid2019}%
  \BibitemOpen
  \bibfield  {author} {\bibinfo {author} {\bibfnamefont {K.~M.}\ \bibnamefont
  {Reid}}, \bibinfo {author} {\bibfnamefont {H.~D.}\ \bibnamefont {Pandey}}, \
  and\ \bibinfo {author} {\bibfnamefont {D.~M.}\ \bibnamefont {Leitner}},\
  }\href {\doibase 10.1021/acs.jpcc.8b11640} {\bibfield  {journal} {\bibinfo
  {journal} {J. Phys. Chem. C}\ }\textbf {\bibinfo {volume} {123}},\ \bibinfo
  {pages} {6256} (\bibinfo {year} {2019})}\BibitemShut {NoStop}%
\bibitem [{\citenamefont {Majumdar}\ \emph {et~al.}(2015)\citenamefont
  {Majumdar}, \citenamefont {Sierra-Suarez}, \citenamefont {Schiffres},
  \citenamefont {Ong}, \citenamefont {Higgs}, \citenamefont {McGaughey},\ and\
  \citenamefont {Malen}}]{Majumdar2015}%
  \BibitemOpen
  \bibfield  {author} {\bibinfo {author} {\bibfnamefont {S.}~\bibnamefont
  {Majumdar}}, \bibinfo {author} {\bibfnamefont {J.~A.}\ \bibnamefont
  {Sierra-Suarez}}, \bibinfo {author} {\bibfnamefont {S.~N.}\ \bibnamefont
  {Schiffres}}, \bibinfo {author} {\bibfnamefont {W.~L.}\ \bibnamefont {Ong}},
  \bibinfo {author} {\bibfnamefont {C.~F.}\ \bibnamefont {Higgs}}, \bibinfo
  {author} {\bibfnamefont {A.~J.}\ \bibnamefont {McGaughey}}, \ and\ \bibinfo
  {author} {\bibfnamefont {J.~A.}\ \bibnamefont {Malen}},\ }\href {\doibase
  10.1021/nl504844d} {\bibfield  {journal} {\bibinfo  {journal} {Nano Letters}\
  }\textbf {\bibinfo {volume} {15}},\ \bibinfo {pages} {2985} (\bibinfo {year}
  {2015})}\BibitemShut {NoStop}%
\bibitem [{\citenamefont {Losego}\ \emph {et~al.}(2012)\citenamefont {Losego},
  \citenamefont {Grady}, \citenamefont {Sottos}, \citenamefont {Cahill},\ and\
  \citenamefont {Braun}}]{Losego2012}%
  \BibitemOpen
  \bibfield  {author} {\bibinfo {author} {\bibfnamefont {M.~D.}\ \bibnamefont
  {Losego}}, \bibinfo {author} {\bibfnamefont {M.~E.}\ \bibnamefont {Grady}},
  \bibinfo {author} {\bibfnamefont {N.~R.}\ \bibnamefont {Sottos}}, \bibinfo
  {author} {\bibfnamefont {D.~G.}\ \bibnamefont {Cahill}}, \ and\ \bibinfo
  {author} {\bibfnamefont {P.~V.}\ \bibnamefont {Braun}},\ }\href {\doibase
  10.1038/nmat3303} {\bibfield  {journal} {\bibinfo  {journal} {Nat. Mater.}\
  }\textbf {\bibinfo {volume} {11}},\ \bibinfo {pages} {502} (\bibinfo {year}
  {2012})}\BibitemShut {NoStop}%
\bibitem [{\citenamefont {Hu}\ \emph {et~al.}(2010)\citenamefont {Hu},
  \citenamefont {Zhang}, \citenamefont {Hu}, \citenamefont {Wang},
  \citenamefont {Li},\ and\ \citenamefont {Keblinski}}]{Hu2010}%
  \BibitemOpen
  \bibfield  {author} {\bibinfo {author} {\bibfnamefont {L.}~\bibnamefont
  {Hu}}, \bibinfo {author} {\bibfnamefont {L.}~\bibnamefont {Zhang}}, \bibinfo
  {author} {\bibfnamefont {M.}~\bibnamefont {Hu}}, \bibinfo {author}
  {\bibfnamefont {J.~S.}\ \bibnamefont {Wang}}, \bibinfo {author}
  {\bibfnamefont {B.}~\bibnamefont {Li}}, \ and\ \bibinfo {author}
  {\bibfnamefont {P.}~\bibnamefont {Keblinski}},\ }\href {\doibase
  10.1103/PhysRevB.81.235427} {\bibfield  {journal} {\bibinfo  {journal} {Phys.
  Rev. B}\ }\textbf {\bibinfo {volume} {81}},\ \bibinfo {pages} {235427}
  (\bibinfo {year} {2010})}\BibitemShut {NoStop}%
\bibitem [{\citenamefont {Ulman}(1996)}]{Ulman1996}%
  \BibitemOpen
  \bibfield  {author} {\bibinfo {author} {\bibfnamefont {A.}~\bibnamefont
  {Ulman}},\ }\href {https://pubs.acs.org/sharingguidelines} {\bibfield
  {journal} {\bibinfo  {journal} {Chem. Rev.}\ }\textbf {\bibinfo {volume}
  {96}},\ \bibinfo {pages} {1533} (\bibinfo {year} {1996})}\BibitemShut
  {NoStop}%
\bibitem [{\citenamefont {Singh}, \citenamefont {Seong},\ and\ \citenamefont
  {Sinha}(2013)}]{Singh2013}%
  \BibitemOpen
  \bibfield  {author} {\bibinfo {author} {\bibfnamefont {P.}~\bibnamefont
  {Singh}}, \bibinfo {author} {\bibfnamefont {M.}~\bibnamefont {Seong}}, \ and\
  \bibinfo {author} {\bibfnamefont {S.}~\bibnamefont {Sinha}},\ }\href
  {\doibase 10.1063/1.4804383} {\bibfield  {journal} {\bibinfo  {journal}
  {Appl. Phys. Lett.}\ }\textbf {\bibinfo {volume} {102}},\ \bibinfo {pages}
  {181906} (\bibinfo {year} {2013})}\BibitemShut {NoStop}%
\bibitem [{\citenamefont {Kl{\"{o}}ckner}\ and\ \citenamefont
  {Pauly}(2019)}]{Klockner2019}%
  \BibitemOpen
  \bibfield  {author} {\bibinfo {author} {\bibfnamefont {J.~C.}\ \bibnamefont
  {Kl{\"{o}}ckner}}\ and\ \bibinfo {author} {\bibfnamefont {F.}~\bibnamefont
  {Pauly}},\ }\href {http://arxiv.org/abs/1910.02443} {\bibfield  {journal}
  {\bibinfo  {journal} {arXiv:1910.02443}\ ,\ \bibinfo {pages} {1}} (\bibinfo
  {year} {2019})},\ \Eprint {http://arxiv.org/abs/1910.02443}
  {arXiv:1910.02443} \BibitemShut {NoStop}%
\bibitem [{\citenamefont {Wu}\ \emph {et~al.}(2020)\citenamefont {Wu},
  \citenamefont {Bates}, \citenamefont {Sangtarash}, \citenamefont {Ferri},
  \citenamefont {Thomas}, \citenamefont {Higgins}, \citenamefont {Robertson},
  \citenamefont {Nichols}, \citenamefont {Sadeghi},\ and\ \citenamefont
  {Vezzoli}}]{Wu2020}%
  \BibitemOpen
  \bibfield  {author} {\bibinfo {author} {\bibfnamefont {C.}~\bibnamefont
  {Wu}}, \bibinfo {author} {\bibfnamefont {D.}~\bibnamefont {Bates}}, \bibinfo
  {author} {\bibfnamefont {S.}~\bibnamefont {Sangtarash}}, \bibinfo {author}
  {\bibfnamefont {N.}~\bibnamefont {Ferri}}, \bibinfo {author} {\bibfnamefont
  {A.}~\bibnamefont {Thomas}}, \bibinfo {author} {\bibfnamefont {S.~J.}\
  \bibnamefont {Higgins}}, \bibinfo {author} {\bibfnamefont {C.~M.}\
  \bibnamefont {Robertson}}, \bibinfo {author} {\bibfnamefont {R.~J.}\
  \bibnamefont {Nichols}}, \bibinfo {author} {\bibfnamefont {H.}~\bibnamefont
  {Sadeghi}}, \ and\ \bibinfo {author} {\bibfnamefont {A.}~\bibnamefont
  {Vezzoli}},\ }\href {\doibase 10.1021/acs.nanolett.0c02815} {\bibfield
  {journal} {\bibinfo  {journal} {Nano Lett.}\ }\textbf {\bibinfo {volume}
  {20}},\ \bibinfo {pages} {7980} (\bibinfo {year} {2020})}\BibitemShut
  {NoStop}%
\bibitem [{\citenamefont {Fang}\ \emph {et~al.}(2020)\citenamefont {Fang},
  \citenamefont {Qian}, \citenamefont {Zhao}, \citenamefont {Li},\ and\
  \citenamefont {Gu}}]{Fang2020}%
  \BibitemOpen
  \bibfield  {author} {\bibinfo {author} {\bibfnamefont {J.}~\bibnamefont
  {Fang}}, \bibinfo {author} {\bibfnamefont {X.}~\bibnamefont {Qian}}, \bibinfo
  {author} {\bibfnamefont {C.~Y.}\ \bibnamefont {Zhao}}, \bibinfo {author}
  {\bibfnamefont {B.}~\bibnamefont {Li}}, \ and\ \bibinfo {author}
  {\bibfnamefont {X.}~\bibnamefont {Gu}},\ }\href {\doibase
  10.1103/PhysRevE.101.022133} {\bibfield  {journal} {\bibinfo  {journal}
  {Phys. Rev. E}\ }\textbf {\bibinfo {volume} {101}},\ \bibinfo {pages}
  {022133} (\bibinfo {year} {2020})}\BibitemShut {NoStop}%
\bibitem [{\citenamefont {Hopkins}\ \emph {et~al.}(2012)\citenamefont
  {Hopkins}, \citenamefont {Baraket}, \citenamefont {Barnat}, \citenamefont
  {Beechem}, \citenamefont {Kearney}, \citenamefont {Duda}, \citenamefont
  {Robinson},\ and\ \citenamefont {Walton}}]{Hopkins2012}%
  \BibitemOpen
  \bibfield  {author} {\bibinfo {author} {\bibfnamefont {P.~E.}\ \bibnamefont
  {Hopkins}}, \bibinfo {author} {\bibfnamefont {M.}~\bibnamefont {Baraket}},
  \bibinfo {author} {\bibfnamefont {E.~V.}\ \bibnamefont {Barnat}}, \bibinfo
  {author} {\bibfnamefont {T.~E.}\ \bibnamefont {Beechem}}, \bibinfo {author}
  {\bibfnamefont {S.~P.}\ \bibnamefont {Kearney}}, \bibinfo {author}
  {\bibfnamefont {J.~C.}\ \bibnamefont {Duda}}, \bibinfo {author}
  {\bibfnamefont {J.~T.}\ \bibnamefont {Robinson}}, \ and\ \bibinfo {author}
  {\bibfnamefont {S.~G.}\ \bibnamefont {Walton}},\ }\href {\doibase
  10.1021/nl203060j} {\bibfield  {journal} {\bibinfo  {journal} {Nano Lett.}\
  }\textbf {\bibinfo {volume} {12}},\ \bibinfo {pages} {590} (\bibinfo {year}
  {2012})}\BibitemShut {NoStop}%
\bibitem [{\citenamefont {Towns}\ \emph {et~al.}(2014)\citenamefont {Towns},
  \citenamefont {Cockerill}, \citenamefont {Dahan}, \citenamefont {Foster},
  \citenamefont {Gaither}, \citenamefont {Grimshaw}, \citenamefont {Hazlewood},
  \citenamefont {Lathrop}, \citenamefont {Lifka}, \citenamefont {Peterson},
  \citenamefont {Roskies}, \citenamefont {Scott},\ and\ \citenamefont
  {Wilkins-Diehr}}]{xsede}%
  \BibitemOpen
  \bibfield  {author} {\bibinfo {author} {\bibfnamefont {J.}~\bibnamefont
  {Towns}}, \bibinfo {author} {\bibfnamefont {T.}~\bibnamefont {Cockerill}},
  \bibinfo {author} {\bibfnamefont {M.}~\bibnamefont {Dahan}}, \bibinfo
  {author} {\bibfnamefont {I.}~\bibnamefont {Foster}}, \bibinfo {author}
  {\bibfnamefont {K.}~\bibnamefont {Gaither}}, \bibinfo {author} {\bibfnamefont
  {A.}~\bibnamefont {Grimshaw}}, \bibinfo {author} {\bibfnamefont
  {V.}~\bibnamefont {Hazlewood}}, \bibinfo {author} {\bibfnamefont
  {S.}~\bibnamefont {Lathrop}}, \bibinfo {author} {\bibfnamefont
  {D.}~\bibnamefont {Lifka}}, \bibinfo {author} {\bibfnamefont {G.~D.}\
  \bibnamefont {Peterson}}, \bibinfo {author} {\bibfnamefont {R.}~\bibnamefont
  {Roskies}}, \bibinfo {author} {\bibfnamefont {J.~R.}\ \bibnamefont {Scott}},
  \ and\ \bibinfo {author} {\bibfnamefont {N.}~\bibnamefont {Wilkins-Diehr}},\
  }\href {\doibase 10.1109/MCSE.2014.80} {\bibfield  {journal} {\bibinfo
  {journal} {Comput. Sci. Eng.}\ }\textbf {\bibinfo {volume} {16}},\ \bibinfo
  {pages} {62} (\bibinfo {year} {2014})}\BibitemShut {NoStop}%
\end{thebibliography}

\end{document}